\documentclass[range]{ar2e_preprint}
\usepackage{ARAstroBib}

\def\simgt{\lower.5ex\hbox{$\; \buildrel > \over \sim \;$}}
\def\simlt{\lower.5ex\hbox{$\; \buildrel < \over \sim \;$}}
\def\farcs{\hbox{$.\!\!^{''}$}}
\newcommand{\kms}          {\mbox{${\rm km~s^{-1}}$}}
\newcommand{\cc}           {\mbox{${\rm cm^{-3}}$}}

\newcommand{\ee}           {\mbox{$^{-2}$}}

\begin{document}

\input epsf.tex    
\input epsf.def   

\input psfig.sty

\jname{..}
\jyear{2010}
\jvol{49}
\ARinfo{1056-8700/97/0610-00}

\title{Protoplanetary Disks and Their Evolution}

\markboth{Williams \& Cieza}{Protoplanetary Disks}

\author{Jonathan P. Williams and Lucas A. Cieza
\affiliation{Institute for Astronomy, University of Hawaii, Honolulu, HI 96822, USA}}


\begin{abstract}
Flattened, rotating disks of cool dust and gas extending for tens
to hundreds of AU are found around almost all low mass stars
shortly after their birth.
These disks generally persist for several Myr, during which time
some material accretes onto the star, some is lost through outflows
and photoevaporation, and some condenses into centimeter- and
larger-sized bodies or planetesimals.
Through observations mainly at infrared through millimeter wavelengths,
we can determine how common disks are at different ages, measure
basic properties including mass, size, structure, and composition,
and follow their varied evolutionary pathways.
In this way, we see the first steps toward exoplanet formation and
learn about the origins of the Solar System.
This review addresses observations of the outer parts, beyond 1 AU,
of protoplanetary disks with a focus on recent infrared and (sub-)millimeter
results and an eye to the promise of new facilities in the immediate future.
\end{abstract}

\maketitle
\section{INTRODUCTION}
\label{sec:introduction}
Circumstellar disks are an inevitable consequence of angular momentum
conservation during the formation of a star through gravitational collapse.
Initially disks rapidly funnel material onto the star but,
as the surrounding molecular core is used up or otherwise disperses,
the accretion rate decreases and a small amount of material persists.
That these disks can be considered protoplanetary is apparent not only
through the geometry of the Solar System but also the high detection
rate of exoplanets.

Because disks exhibit a range of temperatures -- hot near the star,
cooler father away -- they radiate strongly at a range of wavelengths
from microns to millimeters.
They can therefore be observed with infrared and radio telescopes,
and the mapping of wavelength to radius allows detailed models of their
structure to be determined purely from unresolved photometry.
Furthermore, their longevity, relative to the natal core,
allows their properties to be studied in relation to
the optically visible protostar.

Internal friction, or viscosity, within the disk drives continued
accretion onto the star. To preserve angular momentum, some material
is lost through outflows and the disk may gradually spread out with time.
Its structure may also be
strongly affected by photoevaporation, both from the central star
and external stars, and by the agglomeration of dust grains well beyond
the typical sizes found in the interstellar medium including, ultimately,
into planetesimals large enough to gravitationally perturb the disk.
The various evolutionary pathways lead to inner holes and gaps that
reveal themselves through a relative decrement in flux over a
narrow range of wavelengths and which may also be imaged directly
at sufficiently high resolution.

The Infrared Astronomical Satellite (\emph{IRAS}) opened up the infrared
sky and allowed the first statistical studies of disk occurrence to be made
\citep{1989AJ.....97.1451S}.
Shortly thereafter the first sensitive detectors at millimeter wavelengths
showed that many disks contained large dust grains \citep{1989ApJ...340L..69W}
with enough material to form planetary systems on the scale of our own
\citep{1990AJ.....99..924B}.
Interferometry at these long wavelengths provided the ability to resolve
the rotation in the disks \citep{1987ApJ...323..294S},
but unequivocal evidence for their flattened morphology actually came in
the optical, with the Hubble Space Telescope (\emph{Hubble}),
through exquisite images of disk shadows against a bright
nebular background \citep{1994ApJ...436..194O}.
The pace of discoveries has accelerated in the past decade due
to increases in sensitivity, resolution, and wavelength coverage.
The Infrared Space Observatory (\emph{ISO}) and,
in particular, the Spitzer Space Telescope (\emph{Spitzer})
have greatly expanded the known disk inventory in terms of central
stellar mass, age, environment, and evolutionary state.
Interferometry has expanded to longer baselines and shorter wavelengths,
including into the sub-millimeter regime with the Submillimeter Array
(\emph{SMA}),
providing the ability to map fainter structures in greater detail.
The potential to address fundamental questions in protoplanetary disk
studies provided significant motivation for the development of major
new facilities including the
Herschel Space Observatory (\emph{Herschel})
and Atacama Large Millimeter/submillimeter Array (\emph{ALMA}).

The rather short history of protoplanetary disk research may be followed
in the regular Protostars and Planets series
\citep{1978prpl.conf.....G,
1985prpl.conf.....B,
1993prpl.conf.....L,
2000prpl.conf.....M,
2007prpl.conf.....R}.
There are also several related reviews that have been recently written
for this series including the inner disk
\citep{2010ARA&A..48..205D},
debris disks \citep{2008ARA&A..46..339W},
and dynamical processes \citep{2010arXiv1011.1496A}.

This review focuses on the properties and evolution of the outer
parts of protoplanetary disks around low mass stars
as determined principally from
observations at mid-infrared to millimeter wavelengths.
After briefly describing the classification of young stellar
objects in \S\ref{sec:classification}, we begin by discussing
the formation of disks in \S\ref{sec:disk_formation}
and their basic properties when the central star first becomes
optically revealed in \S\ref{sec:disk_properties}.
The second half concerns itself with the temporal properties of disks.
We discuss their lifetimes in \S\ref{sec:disk_lifetimes},
the evidence for, and processes by which, disks evolve
in \S\ref{sec:disk_evolution},
and the properties of disks transitioning into their end state
in \S\ref{sec:transition_disks}.
Each section ends with a short summary of the key points and these,
in turn, are distilled into an overall summary in \S\ref{sec:summary}.
There are many promising avenues for future exploration that new and
planned facilities can address, and we list those that we deem most
exciting in \S\ref{sec:future}.

\section{CLASSIFICATION OF YOUNG STELLAR OBJECTS}
\label{sec:classification}
The process of star and planet formation
begins with the collapse of a molecular core.
The mass is initially all in the core but it is processed through an
accretion disk inward onto the protostar and outward through an
outflow. Ultimately the core is dispersed and the remaining
mass is concentrated in the star.
There is enough terminology associated with this process
to form its own ``diskionary'' \citep{2009arXiv0901.1691E}
and many observational ways to characterize the progression.
The most direct, that of measuring the mass in each component,
is hard and more practical means are generally used, principally measuring
disk accretion signatures in the optical or the distribution of warm
circumstellar material in the infrared.

The infrared based classification dates back to \cite{1984ApJ...287..610L}
who showed that Young Stellar Objects (YSO) in Ophiuchus formed 3 distinct
groups based on whether the emitted energy was rising in the
mid-infrared, declining but with a notable excess over the blackbody
stellar photosphere, or with negligible infrared excess.
This was formalized into 3 classes, I-II-III respectively,
by \cite{1987IAUS..115....1L}
based on the slope of the spectral energy distribution (SED)
between about 2 and $25\,\mu$m,
\begin{equation}
\alpha_{\rm IR}={d\log\nu F_\nu\over d\log\nu}
               ={d\log\lambda F_\lambda\over d\log\lambda}.
\end{equation}
\cite{1994ApJ...434..614G}
subsequently introduced an additional refinement of
``flat-spectrum sources'', intermediate between Class I and II YSOs.
The class sequence was shown to fit naturally into the theoretical framework
of a rotating, collapsing core by \cite{1987ApJ...312..788A}.
As the ability to detect faint millimeter emission improved,
the categorization was extended to an earlier, Class 0,
phase by \cite{1993ApJ...406..122A}
and the decrease in mass of the circumstellar envelope
(material participating in the star/disk formation)
along the sequence was verified
\citep{1994ApJ...420..837A}.

A parallel accretion-based classification exists for the
later optically visible phases: classical and weak-lined T Tauri stars.
These correspond closely (though not exactly) to Class II and III
YSOs respectively. Classical T Tauri stars (CTTS) have strong
H$\alpha$ and UV emission whereas weak-lined T Tauri stars (WTTS)
show no or only very low indications of accretion.
Historically the dividing line between the two was a uniform
H$\alpha$ equivalent width of 10\AA\ but this has been refined
to a stellar mass dependent limit to account for the lower
continuum level in low mass stars
\citep{2003AJ....126.2997B, 2003ApJ...582.1109W}.
The census of Class III sources or WTTS is generally incomplete in
optical and infrared catalogs because they lack strong accretion signatures
and infrared excesses.  Their youth is generally only betrayed
through their location in the Hertzsprung-Russell diagram above the
main sequence or by X-ray activity \citep{1999ARA&A..37..363F}.

It is also important to note that the SED classification does not
give a unique description of the amount and distribution of
circumstellar material.
In particular, YSOs with edge-on disks are highly
extincted and can be mis-interpreted as more embedded,
hence, less evolved, objects.
For example, a Class II YSO viewed at high inclination has a similar
SED to a typical Class I YSO and an edge-on Class I YSO can
have characteristics of a Class 0 YSO \citep{2006ApJS..167..256R}.
These ambiguities highlight the necessity to distinguish between
the physical ``stage'' of YSO evolution for comparison to theory
and its observed SED.
Resolved images, ideally at multiple wavelengths, are required to
fully characterize the evolutionary state of any individual YSO.
Table~1 summarizes the definitions for each class,
where the numerical boundaries for
$\alpha_{\rm IR}$ follow \cite{1994ApJ...434..614G},
including corresponding physical properties (evolutionary stage)
and other observational characteristics.

A disk forms very early on and grows rapidly during the Class 0
collapse phase (see \S\ref{sec:disk_formation}).
On average, the embedded phases through Class I lasts for about
$\sim 0.5$\,Myr \citep{2009ApJS..181..321E}.
The surrounding envelope and strong protostellar outflows can
hinder measurements of disk properties at these early times.
The properties of the disks as their central stars first become optically
visible, i.e. Class II YSOs, are discussed in \S\ref{sec:disk_properties}.
The median disk lifetime after the embedded phase is 2--3\,Myr
but the manner and the rate at which any individual star-disk system
evolves vary greatly. These issues are discussed extensively
in \S\S\ref{sec:disk_lifetimes},\,\ref{sec:disk_evolution}.

Large \emph{Spitzer} surveys have mapped about 90\% of all the
star-forming regions within 500\,pc of the Sun
\citep[][and references therein]{2009ApJS..181..321E}
and spectra have been obtained for over 2000 YSOs therein
\citep[e.g.,][]{2007ApJ...659..680K, 2009ApJ...703.1964F, 
2010ApJ...714..778O}.
The precision and wavelength coverage of these observations
show the tremendous diversity of disk SEDs and the inadequacy of
a single parameter, $\alpha_{\rm IR}$, to characterize the full
range, especially as disks dissipate and open up central holes.
These so-called transition disks are the subject of
\S\ref{sec:transition_disks}.

\section{DISK FORMATION}
\label{sec:disk_formation}

The initial collapse of a molecular cloud core is onto a point source
but a disk quickly forms
as more distant material with higher angular momentum falls inward.
The disk extends out to the centrifugal radius, which is expected
to grow rapidly with time, $R(t) \propto \Omega^2 t^3$,
where $\Omega$ is the angular rotation rate of the core
\citep{1984ApJ...286..529T}.
Disks should evolve rapidly, therefore,
and their final size and mass depends sensitively on the
infall time ($t^3$) and the core properties ($\Omega^2$).
\cite{1998ApJ...509..229B} notes that magnetized collapsing cores may not
be in rigid rotation and that the radius may grow only linearly with time.
In any case, given the wide range of core rotation rates
\citep{1993ApJ...406..528G}
and likely variation in infall duration, we should not be surprised
by an inherent and large diversity in initial disk sizes and masses.

The role of magnetic fields in core collapse and disk formation
is uncertain. Polarization measurements show magnetic field lines
concentrating at the center of collapsing cloud cores in a
pinched, ``hour glass'' configuration \citep{2006Sci...313..812G}
yet Zeeman observations show that the magnetic field strength is
generally insufficient to support cores against their own gravity
\citep{2009ApJ...692..844C}.
We expect some field lines to be dragged down to disk scales for the
magneto-rotational instability, the most likely mechanism for
disk viscosity at late times, to exist.
Singledish sub-millimeter measurements of polarization in disks
\citep{1999ApJ...525..832T} have not been confirmed with
interferometry, however, thus challenging theoretical understanding
\citep{2009ApJ...704.1204H}.

Numerical models of both magnetic and non-magnetic collapsing molecular
cores show disks form rapidly, within $\sim 10^4$\,yr
\citep{1993ApJ...411..274Y, 2005A&A...442..703H}.
Temperatures are very high in these early stages due to gravitational infall.
As the core material is used up or otherwise dispersed,
the disk cools down and its mass decreases as it accretes onto the star.

Core collapse onto a disk opens up an approximately spherical cavity
in the surrounding envelope of radius $R(t)$ that has been inferred from
the presence of excess mid-infrared emission above that expected from a
more extincted centrally peaked core
\citep{2005ApJ...631L..77J, 2009ApJ...707..103E}.
Although there are many observations of inward motions on core-size
scales \citep[e.g.,][]{2001ApJ...562..770D},
the direct detection of gas flow onto a disk has yet to
be convincingly demonstrated.
\cite{2005ApJ...632..371C} find SO line absorption against the
disk continuum in the IRAS~16293-2422B YSO but the spectral
profile is approximately symmetric about the source velocity 
and cannot be clearly identified as pure infall.
\cite{2007Natur.448.1026W} detect many mid-infrared lines of H$_2$O toward
NGC1333-IRAS4B, which they model as arising from a dense, warm, and
compact region.
They attribute this to shocked gas from an envelope onto a disk surface.
However, \cite{2010ApJ...710L..72J} mapped the 1.5\,mm
$3_{1,3}-2_{2,0}$ transition of H$_2^{18}$O from the ground
and find the water emission is quiescent and follows the disk rotation.

Imaging embedded disks requires long wavelengths to see through the
envelopes and arcsecond or higher resolution to match the disk sizes.
Millimeter inteferometers meet these requirements and also filter
out extended structures so that the emission from the compact disk
dominates on long baselines, $\simgt 50{\rm k}\lambda$
\citep[e.g.,][]{1990ApJ...355..635K, 2000MNRAS.319..154B,
2000ApJ...529..477L, 2005ApJ...632..973J}.
A 1.1\,mm continuum survey of 20 embedded YSOs by \cite{2009A&A...507..861J}
shows that the disk flux is an average of four times higher in the embedded
Class 0 phase than in Class I sources but, after allowing for
higher temperatures due to greater accretion heating,
the inferred disk masses show no significant dependence on
evolutionary state. Masses in both Class 0 and I sources range
from $\sim 0.02-0.1\,M_\odot$ with a median $0.04\,M_\odot$.
This is quantitatively similar to the results from
2.7\,mm observations of 6 sources by \cite{2003ApJ...592..255L}.

The lack of dependence of disk mass on evolutionary state from Class 0
to Class I is contrary to theoretical expectations of steady disk growth as
outlined above, whether the core is rigidly or differentially rotating.
Rather, it indicates that disks form quickly and that the flow of material
from the envelope, which declines in mass by almost an order of magnitude
between these two classes \citep{2003ApJS..145..111Y},
is rapidly transported through the disk.

A likely cause of the rapid transport at these early stages
is disk instability.
\cite{1994ApJ...436..335L} first suggested that disks would be
gravitationally unstable during the early formation stages due to
the relatively high mass fraction in the disk
versus that accreted onto the protostar. The instabilities would lead to
sporadic bursts of high accretion \citep{2009ApJ...701..620Z}
and prevent the disk mass from growing faster than the star
\citep{2010ApJ...719.1896V}.
Many protostars have been observed to undergo short-lived bursts of
activity due to high accretion and these events are named after the
first such identified case, FU Orionis \citep{1977ApJ...217..693H}.
\cite{1996ARA&A..34..207H} estimate that a typical low mass star
may have an average of about 10 such outbursts during its formation.

Additional supporting observational evidence for ``punctuated
evolution'' at these early stages is found in measurements
of mass infall rates through the different components and the
protostellar luminosity distribution.
\cite{2005ApJ...635..396E} find envelope infall rates are more than an
order of magnitude higher than disk accretion rates in Class I YSOs
suggesting that mass builds up in the disk until a burst event occurs.
It has long been known that Class I YSOs are, on average,
about an order of magnitude less luminous than expected for the
steady release of gravitational energy as the envelope falls onto the
protostar over the lifetime of the embedded phase \citep{1990AJ.....99..869K}.
Based on the statistics from the cores-to-disks \emph{Spitzer} survey of 5
large, nearby molecular clouds, \cite{2009ApJS..181..321E}
concludes that, on average, a star gains half of its final mass in only
$\sim 7$\% of the $\sim 0.5$\,Myr Class 0 plus I lifetime.
These statistics can be matched by models of episodic mass accretion
from self-gravitating disks \citep{2010ApJ...710..470D}.

Some words of caution are necessary in the interpretation of
the interferometric data, however. The analyses of the continuum
visibilities are based on rather sparse sampling of the Fourier plane
and are subject to confusion with small scale structure in the
envelope \citep{2008ApJ...680..474C}.
A more secure identification of an embedded disk requires
spectral line observations showing rotation.
The emission from the core and any protostellar outflow may still present
significant confusion, although this is somewhat mitigated by observing
species such as HCN and HCO$^+$ that only emit substantially in gas that
is warmer and denser than the outer parts of the envelope.
Recent work demonstrates the possibilities in this area with
indications of Keplerian velocity profiles in moderately
young, Class I sources
\citep{2007A&A...475..915B, 2008A&A...481..141L, 2009A&A...507..861J}.
More sensitive observations of optically thin isotopologues are required
to discern disk kinematics from the core background in the younger,
more embedded Class 0 phase.
Such observations also have the potential to measure the central
protostellar mass and track its growth against the evolution of the
envelope and disk.

\subsection{Section summary} 
\begin{itemize}     
 \item Circumstellar disks form almost immediately
       after a molecular core collapses. Their existence can be
       inferred through both SED modeling and interferometry.
 \item Disk masses do not appear to increase with time during core
       collapse implying a rapid transport onto the star.
 \item The low average luminosity of YSOs and the accretion bursts of
       FU Orionis objects suggest that young disks are (gravitationally)
       unstable.
\end{itemize}

\section{PROPERTIES OF PROTOPLANETARY DISKS}
\label{sec:disk_properties}
The deeply embedded Class 0 and I phases of star formation
only last for a small fraction of a disk lifetime, typically $\sim 0.5$\,Myr
compared to several Myr. By the end of the Class I phase
the envelope has completely dispersed and the star formation process
is effectively over.
The disk now contains only a few percent of the central stellar
mass and can be considered truly protoplanetary, not protostellar.
Although there may be a small amount of accretion of material from the
molecular cloud, the major elements governing the evolution of the disk
at this stage are accretion onto the star, photo-evaporation from local
or external radiation sources, agglomeration into larger bodies,
and dynamical interactions with stellar or substellar companions.

Unless the disk is edge-on to our line of sight, the extinction to
the central protostar is small and its spectral type can be determined
through optical or near-infrared spectroscopy in most cases.
The properties of the disks at this stage marks the baseline for
studies of their evolution and for core accretion models of giant
planet formation.
In this section, we discuss the basic properties of the outer regions
of protoplanetary disks around Class II YSO.

As a fiducial comparison, we use the Minimum Mass Solar Nebula (MMSN),
the lowest mass primordial disk that formed the Solar
System inferred from scaling planetary compositions to cosmic
abundances at each orbital radius
\citep{1970PThPh..44.1580K, 1977MNRAS.180...57W}.
Because of the uncertainty in the composition of the giant planets,
\cite{1977MNRAS.180...57W} quotes a range for the MMSN of $0.01-0.07\,M_\odot$.
Here, we use the minimum of this range, $0.01\,M_\odot\approx 10\,M_{\rm Jup}$,
as an absolute lower mass for the solar nebula out to the 30\,AU orbit
of Neptune.
Extrasolar planetary systems have been found around 10\% of Sun-like
stars in the solar neighborhood, and many planets detected to date
are considerably more massive than Jupiter \citep{2009PASP..121..309J}.
The minimum mass disk required for their formation is proportionally higher.

The extrapolated surface density profile,
which is an observable for resolved disks,
has an approximate power law form, $\Sigma\propto r^{-3/2}$,
although more sophisticated fits based on viscous disk evolution
and planet migration models can be made
\citep{2005ApJ...627L.153D, 2007ApJ...671..878D}.

The properties of the outer parts ($\simgt 1$\,AU) of protoplanetary
disks are mainly inferred from observations at mid-infrared through
millimeter wavelengths although optical and near-infrared scattered
light or silhouette observations can provide important information
on disk radii and dust grain properties
\citep{2008ApJ...673L..63P, 2001Sci...292.1686T}.

\subsection{Mass}
\label{sec:disk_properties_mass}
Disk masses are best determined from (sub-)millimeter wavelength
observations of dust.
The continuum emission is optically thin except in the innermost regions
where column densities are very high. The optical depth is
the integral of the dust opacity, $\kappa_\nu$, times the density, $\rho$,
along the line of sight, $\tau_\nu = \int\rho\kappa_\nu ds = \kappa_\nu\Sigma$,
where $\Sigma$ is the projected surface density.
A commonly used prescription for the dust opacity in disks
at millimeter wavelengths is
\begin{equation}
\label{eq:kappa}
\kappa_\nu=0.1\left({\nu\over 10^{12}\,{\rm Hz}}\right)^\beta~{\rm cm^2\,g}^{-1}
\end{equation}
\citep{1990AJ.....99..924B}.  Both the absolute value and power law index,
$\beta$, are related to the size distribution and composition of the
dust grains \citep{1994A&A...291..943O, 1994ApJ...421..615P}.
The normalization above also implicitly includes a gas-to-dust ratio of 100,
and $\rho$ and $\Sigma$ refer to the total (gas plus dust) density.

For disks, $\beta\approx 1$ so $\kappa(1\,{\rm mm})=0.03~{\rm cm^2\,g}^{-1}$
which implies $\tau(1\,{\rm mm})=1$ at a surface density of
$\Sigma\approx 30~{\rm g\,cm}^{-2}$, corresponding to about 10\,AU
in the MMSN \citep{2005ApJ...627L.153D}
and an angular scale of $0\farcs 07$ in the nearby Taurus star-forming region.
Protoplanetary disks are generally much larger than this and most of the
resolved emission is indeed optically thin.
However, the inner 10\,AU likely constitutes a substantial
fraction of the planet-forming region of a disk. Not only high resolution
but also wavelengths longer than 1\,mm are required to peer into this zone at
the early phases of disk evolution.

For the general question of disk mass measurements on scales much
larger than 10\,AU, we can consider the emission to be optically thin
and therefore directly relate the observed flux, $F_\nu$, to the mass,
\begin{equation}
\label{eq:mass}
M{\rm (gas+dust)} = {F_\nu d^2\over\kappa_\nu B_\nu(T)},
\end{equation}
where $d$ is the distance to the source.
This shows an additional advantage of millimeter wavelengths in
that the Planck function is close to the Rayleigh-Jeans regime,
$B_\nu\approx 2\nu^2 kT/c^2$, and the emission is only linearly,
rather than exponentially, dependent on the dust temperature.

The first large millimeter wavelength surveys to measure disk masses
were carried out at 1.3\,mm using a single element bolometer by
\cite{1990AJ.....99..924B}.  in Taurus-Auriga
and \cite{1994ApJ...420..837A} in Ophiuchus.
Both regions are predominantly forming low mass stars,
$M_* < 1\,M_\odot$ with spectral types K-M.
Using bolometer arrays, which provide better sky subtraction capability,
hence lower noise levels and the ability to detect fainter targets,
\cite{2005ApJ...631.1134A, 2007ApJ...671.1800A}
augmented these surveys in size and expanded the wavelength coverage into
the sub-millimeter regime with observations from $350\,\mu$m to $850\,\mu$m.
The results of this work show that the median mass of Class II YSO disks
is $5\,M_{\rm Jup}$ and the median ratio of disk to stellar mass is 0.9\%.
Figure~\ref{fig:diskmass_histogram} shows the mass distributions are
rather flat by logarithmic mass interval with a steep decline
beyond $\sim 50M_{\rm Jup}$.
The explicit labeling of the gas-to-dust ratio in this figure
mimics an earlier review by \cite{1996Natur.383..139B}
and serves to illustrate the large extrapolation necessary to derive
total disk masses and constrain models of giant planet formation.
This figure also includes the disk census
by \cite{2010ApJ...725..430M} in the more distant but massive star
forming Orion region where photoevaporation is
important (see \S\ref{sec:environment_photoevaporation}).

By modeling the infrared-millimeter SED, \cite{2005ApJ...631.1134A}
showed that the simple mass derivation in equation~\ref{eq:mass}
is indeed reasonably accurate with a characteristic temperature, $T=20$\,K.
This temperature is consistent with CO observations \citep{2004ApJ...616L..11Q}
and theoretical expectations \citep{1997ApJ...490..368C}.
\cite{2005ApJ...631.1134A} also show that for a 100\,AU disk with the mass and surface
density profile of the MMSN, about one third of the emission by mass
is optically thick at $850\,\mu$m.

An additional major uncertainty is the hidden mass in large grains.
Dust grows to much larger
sizes in protoplanetary disks than in the interstellar medium and
these ``pebbles'' or ``snowballs'' can hold considerable mass in a
small solid angle with negligible effect on the SED.
As a rule of thumb, observations at a
wavelength $\lambda$ only constrain the properties of dust grains
out to a maximum size $a_{\rm max}\sim 3\lambda$ (Draine 2006).
Only a few disks have only been detected beyond millimeter wavelengths,
and thus we know little about the general occurrence and distribution of
centimeter and larger sized particles in disks. For a grain size
distribution $n(a)\propto a^{-3.5}$ \citep{1977ApJ...217..425M},
the total mass scales as $a_{\rm max}^{1/2}$
and substantial mass may be undetected.
Detailed modeling by \cite{2001ApJ...553..321D} shows that
the opacity prescription in equation~\ref{eq:kappa}
is valid for $a_{\rm max}\sim 0.3-3$\,mm
but is about a factor of 20 smaller for $a_{\rm max}=1$\,m
implying correspondingly higher masses.
The observational evidence for grain growth from sub-micron to millimeter
sizes and beyond is discussed in \S\ref{sec:disk_evolution_graingrowth}.

The uncertainties in disk mass measurements are large but fortuitously
tend to cancel each other out. We overestimate by assuming an
interstellar gas-to-dust ratio and underestimate by ignoring large bodies.
There are a couple of indications that disk masses as derived from
equations~\ref{eq:kappa} and \ref{eq:mass} are significantly low, however,
which would imply that grain growth is the dominant effect.
First, protoplanetary disk masses derived from (sub-)millimeter photometry
are systematically lower than those estimated from
accretion rates integrated over protostellar ages
\citep{1998ApJ...495..385H, 2007ApJ...671.1800A}.
Second, there are not enough massive disks in nearby star-forming
regions to match the statistics on the incidence of extrasolar
giant planets \citep{2010MNRAS.407.1981G}.

\subsection{Radius}
\label{sec:disk_properties_radius}
Disk sizes are hard to measure because the outer parts are cool and
emit weakly. They are still efficient absorbers, however, and
images of disk silhouettes in Orion against the optically bright
HII region background provide a simple and direct size determination.
\cite{2005A&A...441..195V} measure radii ranging from 50 to 194\,AU
for 22 Orion ``proplyds'' where the disk shadow can be clearly seen
in \emph{Hubble} optical images and an additional two outliers with radii
of 338 and 621\,AU.
The latter object, cataloged as proplyd 114-426, is almost $3''$ in
angular extent, resolved in both radial and vertical dimensions,
and provides perhaps the most famous disk image obtained to date
\citep{1996AJ....111.1977M, 1998ApJ...492L.157M}.
Most of the Orion proplyds are apparent only through the
photoevaporative flows from their surfaces, however,
and \cite{2005A&A...441..195V}
estimate radii for 125 such objects from the size of the ionization front.
Although there is substantial uncertainty associated in this indirect
determination, they infer a median radius $\sim 75$\,AU.
They further note that the sample of \emph{Hubble} proplyds accounts
for only half the stars with known infrared excesses and suggest that
over three quarters of the full sample of Orion disks have radii less
than 75\,AU.

Imaging disks at millimeter wavelengths requires inteferometry
on account of their small angular scales in nearby star-forming regions.
\cite{1996A&A...309..493D} carried out the first large interferometric
survey and resolved many Taurus disks with typical angular sizes of $1-2''$.
A curious problem arose in that the size of
the gas disk, as observed in rotational lines of CO, was found
to significantly exceed the size of the continuum image
\citep{2005A&A...443..945P, 2007A&A...469..213I}.
The differences could not be reconciled for a sharply truncated power
law surface density profile
(i.e., $\Sigma\propto R^{-p}$ for $R\leq R_{out}$,
$\Sigma=0$ for $R>R_{out}$)
without introducing an arbitrary change in either the dust-to-gas ratio
or dust opacity at the radius of the continuum disk.
Similarly, \cite{1996AJ....111.1977M} found that pure power laws or
sharp edges do not match the intensity profiles of Orion proplyd
silhouettes and showed that an exponential decay at the outer boundary
was required. Indeed physical models of viscous accretion disks
\citep[e.g.,][]{1974MNRAS.168..603L, 1998ApJ...495..385H},
predict an exponentially tapered profile of the form,
\begin{equation}
\label{eq:sigma}
\Sigma(R)=(2-\gamma){M_d\over 2\pi R_c^2}\left({R\over R_c}\right)^{-\gamma}\,
          \exp\left[-\left({R\over R_c}\right)^{2-\gamma}\right],
\end{equation}
where $M_d$ is the disk mass, $R_c$ is a characteristic radius,
and $\gamma$ specifies the radial dependence of the disk viscosity,
$\nu\propto R^\gamma$.
\cite{2002ApJ...581..357K} first modeled millimeter wavelength
disk images with this profile, but their data were unable
to differentiate between this and sharply truncated power law fits.
With higher resolution observations,
\cite{2008ApJ...678.1119H} showed that this prescription can naturally
account for the apparent size discrepancy in millimeter imaging as the
the optically thick CO line can be detected further out
along the tapered edge than the optically thin dust continuum.

$R_c$ is a characteristic radius that delineates where the surface density
profile begins to steepen significantly from a power law.
As the disk does not have a sharp edge, a physical disk size must be
specified in terms of an intensity threshold.
A rough estimate of $R_c$ may be obtained by noting that
about 2/3 of the total disk mass (approximately flux) lies within it.
More precisely, \cite{2008ApJ...678.1119H} determine $R_c=30-200$\,AU
for four disks by fitting the profile in equation~\ref{eq:sigma}
to the continuum data.
They compare with sharply truncated power law fits and find
$R_{out}\approx 2R_c$.
The tapered disks extend well beyond $R_{out}$, however,
with midplane densities $n({\rm H_2}) > 10^5$\,\cc\ for $R\simlt 800$\,AU.

By simultaneously fitting $850\,\mu$m interferometric visibilities
and infrared-millimeter SEDs, \cite{2009ApJ...700.1502A, 2010ApJ...723.1241A}
determine a wide range $R_c=14-198$\,AU for 16 disks in Ophiuchus.
Their sample is sufficiently large that they are able to identify
a correlation with disk mass, $M_d\propto R_c^{1.6\pm 0.3}$,
which is likely to be a signature of disk formation physics
rather than an evolutionary sequence. They do not find any
correlation with the central star properties.

\cite{2009ApJ...701..260I} use the same density prescription to fit a
sample of 11 disks mostly in Taurus-Auriga. They use a slightly
different radial normalization that converts to a similar range,
$R_c\simeq 30-230$\,AU, and also shows that the older disks tend
to be larger.
In addition, \cite{2009ApJ...701..698S} compile radii for 16 disks
around low mass stars in Taurus-Auriga based on sharply truncated
power law fits to CO observations.
Because of the different fitting technique, their values,
$R_{out}\sim 100-1100$\,AU, are not directly comparable with
the above $R_c$ measures, but they similarly illustrate the large range
of disk sizes and independence on stellar type.

The specific angular momenta of protoplanetary disks, $j=J/M$,
lies in the range $\log_{10}j\,\rm{[cm}^2\,\rm{s}^{-1}{\rm]}=19.4-20.9$
\citep{2009ApJ...701..260I, 2010ApJ...723.1241A}.
This is comparable to the giant planets in the Solar System,
$\log_{10}j=20.0-20.4$, but significantly lower than the
observed values in molecular cores,
$\log_{10}j=20.8-22.6$ \citep{1993ApJ...406..528G}.
Either the angular momentum in a core is not efficiently transferred
to the disk, or its value is overestimated due to the inherent smoothing
of turbulent fluctuations in radial velocity observations
\citep{2010ApJ...723..425D}.

\subsection{Structure}
\label{sec:disk_properties_structure}
The average radial, vertical, and velocity profiles of relatively bright
and large disks can be determined through resolved observations and careful
modeling of spectral energy distributions.

\subsubsection{Surface density}
\label{sec:disk_properties_surfacedensity}
A resolved image of a disk at millimeter wavelengths provides not only
a measure of its total mass and radius but also the distribution of mass,
or surface density.
Until recently this was characterized as a pure power law,
$\Sigma\propto R^{-p}$, with values of $p$ generally in the range $0-1$
\citep{1996ApJ...464L.169M, 2000ApJ...534L.101W, 1997ApJ...489..917L,
2002ApJ...581..357K, 2007ApJ...659..705A}.

The exponential tapered fits of the form in equation~\ref{eq:sigma}
approximate a power law, $\Sigma\propto R^{-\gamma}$ for $R\ll R_c$, 
but the fitted values of $R_c\simeq 30-200$\,AU correspond to $\simlt 1''$
in all but the closest disks and $\gamma$ is actually determined
largely from the steepness of the exponential taper.
The \cite{2008ApJ...678.1119H} comparison of pure power law versus
exponentially truncated power law fits shows similar indices
but with slightly steeper pure power law fits due to the soft
edge, $\langle p\rangle = 1.2, \langle\gamma\rangle = 0.9$ for
four disks.

From larger samples,
\cite{2009ApJ...700.1502A, 2010ApJ...723.1241A}
find a tight range consistent with
all having the same value $\langle\gamma\rangle = 0.9$.
Using a different modeling technique, however,
\cite{2009ApJ...701..260I} find a very wide range
$\gamma=-0.8$ to 0.8 with mean $\langle\gamma\rangle = 0.1$ in their data.
Negative values of $\gamma$ correspond to decreasing surface densities
for $R<R_c$ which may be an important signature of disk evolution
(see \S\ref{sec:disk_evolution}) but these were also found
in the smaller disks with $R_c < 100$\,AU which are barely resolved
and thus the hardest to characterize.

In general, all results agree that young protoplanetary disks
have flatter central density profiles than the canonical power
law $p=1.5$ MMSN \citep{1977MNRAS.180...57W}. There is even more
uncertainty in the density profile of the MMSN than its mass,
however, and an exponential tapered power law fit by \cite{2005ApJ...627L.153D}
has $\gamma=0.5$. The more relevant comparison is of absolute
values in the planet-forming zone.
\cite{2009ApJ...700.1502A, 2010ApJ...723.1241A} infer surface densities,
$\Sigma\approx 10-100$\,g\,cm\ee\ at 20\,AU,
in their sample of Ophiuchus disks that are in good agreement
with the MMSN (Figure~\ref{fig:sigma_sma}).

Whereas the disk-to-star mass ratio may be very high during
the initial stages of formation (\S\ref{sec:disk_formation}),
the Toomre Q parameter, $Q(R)=c\Omega/\pi G\Sigma$
where $c$ is the sound speed and $\Omega$ the orbital angular
velocity \citep{1964ApJ...139.1217T},
is generally much greater than unity for these Class II YSO disks
implying that they are gravitationally stable at all radii
\citep{2009ApJ...701..260I, 2010ApJ...723.1241A}.
The one possible exception is DG Tau, a young protostar with
by far the most massive disk in the two surveys.

The inferred surface densities are increasingly uncertain closer to
the star due to the limited resolution of the observations, $\simgt 20$\,AU,
and also because the emission is becoming optically thick.
As the discrepancy between
\cite{2009ApJ...700.1502A, 2010ApJ...723.1241A}
and \cite{2009ApJ...701..260I} shows,
different data and fits extrapolate to very different surface densities
at the $5-10$\,AU orbital radii of Jupiter and Saturn.
This is not only a critical region
for understanding planet formation but also for disk structure.
The density is so high here that the midplane is shielded
from radiation and cosmic rays.
Unless radioactive elements are sufficiently abundant
\citep{2009ApJ...703.2152T},
the ionization fraction may be so low that matter is decoupled from
the magnetic field and avoids the magneto-rotational instability
(see \S\ref{sec:disk_evolution_viscous}).
Current observations cannot address these issues, and they will remain
a challenge until the combined requirements of high sensitivity
and high resolution at long wavelengths are met
\citep{2004NewAR..48.1363W}.
Note, however, that the uncertainty in the surface density is outweighed
by the Keplerian increase in rotation rate and Class II YSO disks
are gravitationally stable in their inner regions.

\subsubsection{Scale height}
\label{sec:disk_properties_scaleheight}
Protoplanetary disks are flared with a vertical scale height
that increases with radius.
\cite{1987ApJ...323..714K}
first suggested the possibility of flaring
based on the large far-infrared excesses detected by \emph{IRAS} that could
not be explained by a spatially flat disk.
Direct evidence for flared disks can be found in the
beautiful \emph{Hubble} images of Taurus disk silhouettes against the
scattered light of the central star
\citep{1996ApJ...473..437B, 1998ApJ...502L..65S, 1999AJ....117.1490P}
and in Orion against the nebular background
\citep{2005AJ....129..382S}.

Characterizing the disk scale height is essential for modeling
the thermal, ionization and chemical structure of disks and,
thence, for interpreting atomic and molecular line observations.
It is also important for understanding disk evolution as the
outer parts, with their low densities and large scale heights,
are particularly susceptible to photoevaporative losses.

For an azimuthally symmetric disk in hydrostatic equilibrium,
the density is a function of both radius, $R$, and vertical height, $Z$,
\begin{equation}
\label{eq:rho}
\rho(R,Z)={\Sigma(R)\over \sqrt{2\pi}H}\,\exp\left(-{Z^2\over 2H^2}\right),
\end{equation}
where $\Sigma(R)$ is defined in equation~\ref{eq:sigma},
and $H(R)$ is the scale height which depends on the
competition between disk thermal pressure and the vertical
component of stellar gravity.
The disk temperature, in turn, depends on the amount of stellar
radiation impacting the disk and therefore on its geometry.
\cite{1997ApJ...490..368C} provide an elegant analytic solution
to these coupled equations by approximating the radiative transfer
with a hot surface layer that absorbs and reprocesses the starlight,
which then heats a flared interior.
They find an approximate power law dependence, $H\propto R^h$,
with $h\approx 1.3-1.5$.
\cite{1998ApJ...500..411D} and \cite{2002A&A...389..464D}
iterate numerical solutions to the vertical disk
structure and find similar results.

Observed disk SEDs have less mid-infrared emission than expected for this
degree of flaring due to settling of dust grains toward the
midplane \citep{2001ApJ...547.1077C, 1999ApJ...527..893D}.
This flattening of the dust disk relative to the gas occurs
very early in the evolution of a disk
and is discussed further in \S\ref{sec:disk_evolution}.

\subsubsection{Velocity}
\label{sec:disk_properties_velocity}
Disk masses at the Class II stage and beyond are a small fraction of the
central stellar mass. Their motions are therefore expected to be Keplerian.
Velocity profiles have been measured for only a handful of disks.
The difficulty lies in finding disks that are sufficiently young and
therefore bright enough to image in a spectral line, generally a
millimeter wavelength rotational line of CO or isotopologue,
yet not contaminated by emission from the residual envelope or
neighboring cloud.

\cite{1993Icar..106....2K}
resolved the $^{13}{\rm CO}~J=2-1$ line in the
relatively large GM Aur disk and found the velocity profile was
consistent with being Keplerian.
Similarly, \cite{1994A&A...286..149D} imaged the large circumbinary disk
around GG Tau in $^{13}{\rm CO}~J=1-0$ and used a Keplerian velocity
profile to determine a total stellar mass of $1.2\,M_\odot$.
With improvements in instrumentation, leading to higher frequency and
higher resolution observations, more studies followed shortly thereafter
\citep[e.g.,][]{1997Natur.388..555M, 1998A&A...332..867D}.

\cite{1998A&A...339..467G}
modeled the CO $J=1-0$ channel maps (images
at different velocities) of DM Tau to show that a Keplerian velocity
profile was not only consistent but was the best fit.
This was further exploited in the survey by
\cite{2000ApJ...545.1034S} who determined dynamical masses
for 9 systems and tested models of protostellar evolution.
Schaefer et al. (2009) extended this work to later spectral types.
As discussed in \S\ref{sec:disk_formation}, Lommen and Brinch
have used the same technique to derive protostellar masses
at even earlier times in Class 0 and I YSOs.
Spectral line observations also determine disk inclinations to
our line of sight with high accuracy
\citep[e.g.,][]{2004ApJ...616L..11Q},
which is necessary for detailed modeling and characterization
of other disk properties.

Many of the aforementioned studies also note that the line emission
is well fit by only rotational plus thermal broadening.
High spectral resolution observations of the TW Hydra and HD163296
disks by \cite{2010arXiv1011.3826H} show that the turbulent
component is subsonic, $\leq 10$\% and 40\%, of the sound speed respectively
at the $\sim 100$\,AU scales of their observations.
Such a low level of turbulent stirring provides ideal conditions for
grain settling and growth, planetesimal congregation,
and protoplanetary accretion.
Their data also show near-perfect Keplerian rotational profiles
confirming that disk self-gravity does not appear to be significant
in these objects (Figure~\ref{fig:momentmaps}).

\subsubsection{Departures from azimuthal symmetry}
\label{sec:disk_properties_nonazimuthal}
The discussion in this section has so far considered only the radial
or vertical variation of disk properties on account of the dominant
gravity of the central star.
Azimuthal variations are of great interest, however, because they would
signpost an additional influence, potentially disk self-gravity or
proto-planets.

Disks are tiny compared to the distances between stars
even in dense clusters and disruptive encounters are highly unlikely.
For typical relative motions of 1\,\kms, the stellar density must
be greater than $\sim 10^6$ stars\,pc$^{-3}$ for an encounter
within 1000\,AU of a disk within 1 Myr.
Almost all disks that have been studied in detail are nearby and are
either isolated or in low mass star-forming regions where the stellar
densities are orders of magnitude lower than this.
Consequently, the likelihood that any azimuthal distortions in a
young disk are due to an encounter with a nearby star is very low.

The surface density measurements described in
\S\ref{sec:disk_properties_surfacedensity}
and Keplerian velocity profiles in \S\ref{sec:disk_properties_velocity}
show that disk self-gravity is negligible by the time of the Class II phase.
As discussed in \S\ref{sec:disk_formation}, however, younger disks in formation
probably have much higher disk-to-star mass ratios.
The disk around the partially embedded Herbig Ae star AB Aur
may be an example of an unstable disk.
Optical and near-infrared coronagraphic imaging by
\cite{1999ApJ...523L.151G} and \cite{2004ApJ...605L..53F}
show spiral features in scattered light.
Lower resolution millimeter interferometry confirms the larger
scale asymmetries in the system \citep{2005ApJ...622L.133C}
and deviations from Keplerian rotation that may indicate streaming
motions along the arms or an unseen companion
\citep{2005A&A...443..945P, 2006ApJ...645.1297L}.

The gravitational effect of planets on debris disks has been diagnosed
in Fomalhaut \citep{2009ApJ...693..734C},
postulated in $\beta$ Pic \citep{1997MNRAS.292..896M},
and suspected in others \citep[and references therein]{2008ARA&A..46..339W}.
Younger disks are further away than these systems and,
the intriguing case of AB Aur notwithstanding,
the resolution is not yet sufficient to search for such effects,
although this is an exciting prospect for the future
\citep{2005ApJ...619.1114W, 2006ApJ...647.1426N}.

\subsection{Composition}
As with the interstellar medium, the circumstellar medium is composed
of dust and gas. There is substantial processing of both components,
however, as a disk forms and evolves.

\label{sec:disk_properties_composition}
\subsubsection{Dust}
\label{sec:disk_properties_dust}
Dust dominates the opacity of protoplanetary disks and is the
raw material for planetesimals.
In the diffuse interstellar medium, dust is mainly composed of
silicates with sizes $r\simlt 0.1\,\mu$m with an admixture of graphite
grains and polycyclic aromatic hydrocarbons \citep{2003ARA&A..41..241D}.
During the passage through cold dark clouds to protoplanetary disks,
molecules freeze out from the gas phase onto dust grain surfaces producing
icy mantles \citep{2007ARA&A..45..339B}
and the grains collisionally agglomerate \citep{2008ARA&A..46...21B}.

Silicates are readily identified through broad spectral bands
at 10 and $18\,\mu$m \citep{2010ARA&A..48...21H}.
Polycyclic aromatic hydrocarbons also have several mid-infrared features
in this wavelength range \citep{2008ARA&A..46..289T}.
About 2000 protoplanetary disks were observed with the
Spitzer Infrared Spectrometer (IRS) providing a wealth of
information on the composition, size distribution, and evolution of the dust.
We defer discussion to \S\ref{sec:disk_evolution_graingrowth}
and note here only that the dust in disks has been significantly
processed via thermal annealing and agglomeration to have a much
higher crystallinity fraction and larger sizes relative to the
interstellar medium.

\subsubsection{Gas}
\label{sec:disk_properties_gas}
Gas amounts to 99\% of the total mass of the interstellar medium
and the same is true, at least initially, for disks.
Its detection, however, presents a tougher observational
challenge than the minority dust component because it emits only at
specific wavelengths and therefore requires high resolution spectroscopy.
Moreover, at wavelengths where the dust emission is optically thick,
a line can only be seen if its excitation temperature differs from
that of the dust.

Disk accretion, which can be measured through recombination line emission
or excess hot continuum emission, provides unambiguous evidence for
the presence of gas. However, the precise amount or physical
conditions of the gas in the disk cannot be determined from
these diagnostics \citep{2009apsf.book.....H}.

At the typical high densities and low temperatures of the bulk
of a protoplanetary disk, the primary gaseous species is H$_2$.
Fluorescent electronic transitions of H$_2$ in the UV arise from
hot gas very close to the protostar \citep{2006ApJS..165..256H}.
Near-infrared ro-vibrational lines also probe the inner disk, $\simlt 1$\,AU.
Because the focus of this review is on the outer disk, we discuss
mid-infrared observations of H$_2$ only and refer the reader to
\cite{2007prpl.conf..507N} for details on the inner disk gas component.

As a symmetric molecule, H$_2$ has no electric dipole moment and only
magnetic quadrupole transitions are permitted, i.e. $\Delta J=\pm 2$.
These lines lie in the mid-infrared, with energies characteristic of
the temperatures at several AU in a disk,
a potentially very interesting regime for understanding giant planet formation.
\cite{2007ApJ...661L..69B} detected and spectrally resolved 3 lines
(H$_2$ S(1), S(2), S(4) corresponding to $J=3-1, 4-2, 6-4$ respectively)
in AB~Aur.  The linewidths locate the emitting region to 18\,AU in
radius but the rotational diagram indicates a temperature of 670\,K,
which is much hotter than the dust temperature at this radius.
They infer that the gas is decoupled from the dust and
excited by either protostellar UV or X-rays,
or shocks from the infalling envelope \citep[see also][]{2008ApJ...688.1326B}.
Similarly, \cite{2007ApJ...666L.117M} detect the H$_2$ S(1) line in another
intermediate mass Herbig Ae star, HD97048,
and show that the high line-to-continuum ratio
requires either an enhanced gas-to-dust ratio or an additional
gas heating mechanism such as X-rays or shocks.

In both cases, the inferred gas mass is very small.
For AB Aur, where the multiple lines allow the level
populations to be determined, $M_{\rm H_2}=0.5\,M_\oplus\simeq 0.1$\%
of the total disk mass.
Effectively, only the top of the disk atmosphere is measured
due to the high optical depth of the dust at these wavelengths.
Based on the \cite{1997ApJ...490..368C} two-layer model,
\cite{2008A&A...477..839C} calculate that mid-infrared H$_2$ detections
require gas temperatures twice as high as the dust and a gas-to-dust
ratio greater than 1000. This is possible through dust settling
to the midplane but the low detection rates suggest that
such extreme conditions are rare
\citep{2010A&A...516A.110M, 2007ApJ...665..492L}.

Trace amounts, $M_{\rm HI}\sim 1\,M_\oplus$, of atomic hydrogen should
also be present due to photodissociation of H$_2$ \citep{2007ApJ...660..469K}.
Although beyond the sensitivity of current instruments, it will be
possible to detect the 21\,cm line in many disks with the Square
Kilometer Array \citep{2008PhST..130a4013K}.

Fortunately, there are many other currently observable species present in
disks from ionized and atomic at the disk surface to molecular deeper within.
An interesting case is the optical line of [O I] at 6300\,\AA,
which has been detected in the surface of strongly flared Herbig Ae/Be disks
out to $\sim 100$\,AU \citep{2005A&A...436..209A}.
It originates from the photodissociation of OH which results
in excited neutral atomic oxygen.

\emph{Spitzer} spectroscopy has provided many new results.
[Ne II] is one of the strongest lines and has an origin in YSO jets
\citep{2010A&A...519A.113G}.
For sources without jets, the [Ne II] line is much weaker and traces hot
gas on the disk surface, possibly from a photoevaporate flow
\citep{2009ApJ...702..724P}.
Ro-vibrational lines from the organic molecules, C$_2$H$_2$, HCN, and CO$_2$,
have also been detected by IRS, in absorption through a nearly-edge
on disk by \cite{2006ApJ...636L.145L},
and in emission by \cite{2008Sci...319.1504C}.
Assuming local thermodynamic equilibrium, both studies find molecular
excitation temperatures in the range of $300-700$\,K,
implying a location in the inner $3-6$\,AU of the disks.
Their production is due to a rich gas phase chemistry sustained by
the sublimation of icy grain mantles in a warm molecular region above the
disk midplane, and is strongly dependent on the stellar radiation field.
\cite{2009ApJ...696..143P} find a deficit of HCN in disks around
very low mass stars and brown dwarfs, which they
attribute to a reduced photodissociation rate of N$_2$.
\cite{2010ApJ...720..887P} 
find almost no molecular emission toward 25 Herbig Ae/Be stars
(the exception being one CO$_2$ line).
They suggest either photodissociation, lower gas-to-dust ratios,
or masking by the strong continuum as the cause.

The bulk of the lines in the \cite{2008Sci...319.1504C} IRS spectrum
of AA Tau are rotational lines of OH and H$_2$O. These are important
coolants and also of great astrobiological interest.
Their derived high abundances require vertical and potentially radial
transport of ices from the inner, cooler regions of the disk.
\cite{2008ApJ...676L..49S} found similar results toward an additional
two disks, and the large survey by \cite{2010ApJ...720..887P} shows
that these lines are common in T Tauri disks.
The complex origin of the gas-phase water and its implications for
understanding disk structure and evolution have been recently discussed by
\cite{2009ApJ...704.1471M} and \cite{2006Icar..181..178C}.

There are also many detectable molecular lines in the (sub-)millimeter
wavelength regime. Observations of low energy rotational transitions,
which can be excited even at the low temperatures and densities of the
outer disk, provide the best constraints on the total gas mass in a disk.
CO lines are the strongest on account of its high abundance and large
dipole moment. The dominant source of opacity in this case is not the
dust but the lines themselves.
Interferometric CO images therefore show the radial gas temperature gradient,
$T\propto r^{-q}$, with values of $q\simeq 0.4-0.7$
for both T Tauri and Herbig Ae/Be stars
\citep{1998A&A...339..467G, 2007A&A...467..163P},
which is in agreement with the dust temperature derived from
SED modeling \citep{2005ApJ...631.1134A}.
Multi-transition observations trace different $\tau=1$ surfaces
providing information on the vertical temperature structure.
\cite{2006ApJ...636L.157Q} show that the strong CO 6--5 emission of
the TW Hydra disk
requires a hot surface layer with a gas-dust temperature difference of
up to 30\,K that can be explained by X-ray heating.
Observations of the rarer isotopologues, $^{13}$CO and C$^{18}$O,
probe deeper into the disk and show cooler midplane temperatures,
$\sim 13$\,K, and an order-of magnitude depletion of CO relative
to the interstellar medium \citep{2003A&A...399..773D}.
This is due to CO condensation onto dust grains at a temperature of
about 20\,K \citep{1993ApJ...417..815S}.
Similarly high depletions have been found in other T Tauri disks
\citep[e.g.,][]{2004ApJ...616L..11Q, 2010ApJ...720..480O}.
The situation is quite different for disks around the more luminous
Herbig AeBe stars where the dust is warmer than the CO
condensation temperature throughout.
Within the uncertainties of CO isotopologue abundance and dust
opacity, \cite{2005A&A...443..945P} and \cite{2008A&A...491..219P}
show that the gas-to-dust ratios in the AB\,Aur and HD\,169142 disks
are consistent with the interstellar medium value of 100.

Rotational lines of other molecules further constrain the physical
structure as well as the chemistry of the outer disk.
The first results were obtained using single-dish telescopes,
measuring the globally averaged properties, and showed that
other molecules including HCO$^+$ and H$_2$CO were also strongly
depleted as much as, if not more than, CO in classical T Tauri disks
\citep{1997A&A...317L..55D, 2001A&A...377..566V, 2004A&A...425..955T}.
Photodissociated products such as CN were relatively abundant,
however, which led to the picture of a photon dominated region
at the surface of a warm intermediate layer
with $T\sim 20-40$\,K, $n({\rm H_2})\sim 10^6-10^8$\,\cc,
sitting on a cold, highly depleted midplane.
The details in this picture are being filled in as new lines and
species are detected \citep[e.g.,][]{2010ApJ...714.1511H}.

\cite{2004A&A...425..955T} found a high deuterium fraction in TW Hydra,
similar to values seen in dense cores and attributed to cool
temperatures and high CO depletion \citep{1999ApJ...523L.165C}.
Interferometric imaging resolves the D/H ratio, shows that it
varies from $\sim 0.01$ to $\sim 0.1$ across the disk and
peaks at $\sim 70$\,AU radius \citep{2008ApJ...681.1396Q},
a result with implications for understanding the high D/H ratio in
comets and the delivery of water to Earth \citep{2000ARA&A..38..427E}.
The suite of molecular line maps in Figure~\ref{fig:twhya_chemistry}
directly shows the decrease in DCO$^+$ 3--2 intensity
at the center of the disk where the CO depletion is lower due to
the higher temperatures.
Deuterated species, in particular H$_2$D$^+$, and other molecules
with very low condensation temperatures such as N$_2$H$^+$
should reveal the physical conditions and kinematics of the disk midplane
in the outer disk.
The lines are weak but H$_2$D$^+$ has been detected in one,
possibly two, sources by \cite{2004ApJ...607L..51C}
and N$_2$H$^+$ has been detected in several disks
\citep{2003ApJ...597..986Q, 2007A&A...464..615D, 2010ApJ...720..480O}.
The cool temperatures of T Tauri star disks lead to CO freeze out
and a richer chemistry that is absent in warmer
Herbig AeBe disks \citep{2010ApJ...720..480O}.
The reader is referred to \cite{2009arXiv0908.3708B}
for a more exhaustive discussion of disk chemistry.

The full analysis of the many species and lines now detected in
protoplanetary disks requires sophisticated models to handle the
physical, thermal, and chemical structure and the feedback of
each of these on each other \citep[e.g.,][]{2009A&A...501..383W}.
The complexity is such that, for T Tauri stars at least,
there is no simple prescription for the gas mass.
However, these observations provide substantial
information about the disk structure and complement the broadband
dust measurements. Disks have a dynamic atmosphere above a cool midplane:
the temperature increases with height, the phase changes from molecular
to atomic to ionized, and dust grains undergo sublimation and condensation
of icy mantles as they move throughout.

Finally,
as this review is being written, the first results are coming in from
\emph{Herschel} and its far-infrared spectrometers.
The [OI]\,$63\,\mu$m line is generally the strongest line in this regime
and widely detected \citep{2010A&A...518L.127M}.
Combined with ground based CO measurements, it provides strong
constraints on the gas mass \citep{2010A&A...518L.125T}.
Multiple transitions of this and other species can be detected
in some sources and allow the excitation conditions of the gas
to be determined \citep{2010A&A...518L.129S}.
However, the weakness of H$_2$O lines at sub-millimeter wavelengths
indicates that almost all the water is frozen onto icy mantles around
large dust grains that have
settled to the midplane \citep{2010A&A...521L..33B}.
In \emph{Herschel}'s 3 year lifetime, we expect to learn much more about
the gaseous content of disks in the giant planet-forming zone.

\subsection{Dependence on stellar mass}
\label{sec:disk_properties_mass_dependence}
Disks are observed around a wide range of stars from very
low mass to intermediate mass Herbig Ae/Be stars.
Infrared excesses from disks around brown dwarfs are also
detected at a similar frequency to stars \citep{2005ApJ...631L..69L}
and masses for a handful have been measured from
millimeter wavelength measurements \citep{2006ApJ...645.1498S}.
The relationship between disk and stellar properties informs
models of disk formation and evolution.

Higher mass stars require more material to pass through a disk
and we might therefore expect to see a positive correlation
between disk mass and stellar mass \citep[e.g.,][]{2000prpl.conf..559N}.
Figure~\ref{fig:mass_by_star} compiles (sub-)millimeter measurements
of Class II YSO disks from the literature
\citep{2006ApJ...645.1498S, 2009A&A...497..117A, 2000prpl.conf..559N, 2004A&A...422..621A, 2000ApJ...529..391M}
and shows that there is a large scatter,
$\sim 0.5$\,dex, but confirms that disk masses tend to be lower around
low mass stars such that the ratio, $M_d/M_*\sim 0.01$.

The relationship breaks down for the most massive stars. There is no
(sub-)millimeter detection of a disk around an optically visible O star.
The limits on the sensitive Orion proplyd survey provide the most
stringent constraints to date,,
$M_d/M_*\simlt 10^{-4}$ for $M_*\geq 10\,M_\odot$
\citep{2009ApJ...694L..36M}.
This may be due to very high photoevaporation rates so that disks are
not detectable by the time an O star is optically visible or due to an
altogether different star formation mechanism \citep{2007ARA&A..45..481Z}.

Disk size measurements, whether from optical observations of Orion
proplyds \citep{2005A&A...441..195V} or millimeter visibilities of
Ophiuchus disks \citep{2009ApJ...700.1502A, 2010ApJ...723.1241A},
are of a restricted range of stellar masses,
$M_*\simeq 0.3-2\,M_\odot$, and do not follow a clear trend.
Moreover, \cite{2009A&A...508..707P} argue that observations
of Herbig Ae/Be stars at millimeter wavelengths have been biased
toward large, bright disks.
\emph{ALMA} will soon provide the resolution and sensitivity
required to measure disk radii in a more representative sample.

Disk structure does appear to be dependent on stellar mass,
at least for low mass stars.
\cite{2010ApJ...720.1668S} find that relatively blue mid-infrared colors
imply smaller scale heights for disks around very low mass stars than 
low mass stars in the same star-forming region.
It is harder to make a direct comparison to higher mass Herbig Ae/Be stars
in other regions because of the difficulty in distinguishing initial
conditions from evolution due to grain growth and settling.

\subsection{Section summary} 
\begin{itemize}     
 \item Disk masses are best measured from (sub-)millimeter observations
       of thermal continuum emission from dust. Substantial, and uncertain,
       corrections are required for the grain size distribution and gas-to-dust
       ratio.
 \item Assuming an ISM gas-to-dust ratio of 100 and ignoring the mass in
       bodies larger than 1\,mm, the median mass of protoplanetary
       disks around Class II YSOs with spectral types GKM is $5\,M_{\rm Jup}$.
 \item Protoplanetary disks have power law surface density profiles with
       an exponential taper that produces a soft edge anywhere from
       20 to 200\,AU from the star. The power law index is approximately
       $-1$ but is extrapolated and highly uncertain within the central
       $\sim 20$\,AU.
 \item Surface densities are generally below the threshold for gravitational
       instability and velocity profiles are Keplerian.
 \item Mid-infrared through millimeter spectroscopy reveals
       a warm molecular layer with PDR-like chemistry that is strongly
       dependent on the stellar luminosity and a cold midplane
       depleted of molecules.
 \item From brown dwarfs to B stars, disk masses scale with the stellar mass.
       The median ratio is about 1\% but there is a large dispersion,
       $\pm 0.5$\,dex.
\end{itemize}

\section{DISK LIFETIMES}
\label{sec:disk_lifetimes}
One of the most fundamental parameters on disk evolution studies is
the lifetime of the disk itself. This is not only because it reflects the
relevant time scale of the physical processes driving the dissipation
of the disk, but also because it sets a limit on the time available for
planet formation. 

Even though the mass of the gas initially dominates that of the dust
by two orders of magnitude in primordial disks, the dust is much easier
to observe and thus most of the constraints on the lifetimes of
circumstellar disks have been obtained by observing the thermal emission
of the  dust grains. These particles absorb stellar light and reradiate
mostly in the $1\,\mu$m to 1\,mm range.
Because the temperature of the dust decreases with the distance from
the central star, different wavelengths trace different disk radii 
(for a given stellar luminosity).
In what follows, we summarize the observational constraints provided
by near-infrared and mid-infrared observations, mostly from \emph{Spitzer},
which has performed the most sensitive and comprehensive surveys of
disks in star-forming regions to date.
Toward the end of the section, we also briefly discuss current
constraints on gas dispersal time scales.

\subsection{Near-infrared results: the inner disk}
\label{sec:disk_lifetimes_nir}
Because there is a very well established correlation between the presence
of near-infrared excess ($1-5\,\mu$m) and the occurrence of spectroscopic
signatures of accretion \citep{1995ApJ...452..736H},
it is possible to investigate the lifetime of inner accretion disks
($R\simlt 0.1$\,AU) by studying the fraction of stars with near-infrared
excess as a function of stellar age. Early studies of nearby star-forming 
regions found that $60-80$\% of stars younger than 1\,Myr present measurable
near-infrared excesses, and that no more than 10\% of stars older than 10\,Myr
do so \citep[e.g.,][]{1989AJ.....97.1451S}.
Given the large uncertainties associated with model-derived stellar ages,
it has been argued that individual star-forming regions lack the
\emph{intrinsic} age spread necessary to investigate disk lifetimes
from individually derived ages and the apparent age dispersion is mostly
driven by the observational uncertainties \citep{2001AJ....121.1030H}.
However, similar studies based on the disk frequencies in clusters with
a range of \emph{mean} ages \citep{2001ApJ...553L.153H, 2005astro.ph.11083H}
have led to essentially the same result:  the frequency of inner
accretion disks steadily decreases from $<1$ to $\sim 10$\,Myr
(e.g., see Figure~2 in \cite{2008ARA&A..46..339W}).

The inner disk fractions observed in both young stellar clusters and
in the distributed population of pre-main sequence stars in
star-forming regions are consistent with median disk lifetimes of
between 2 and 3 Myr.  The decrease has also been modeled as an
exponential with e-folding time 2.5\,Myr \citep{2009AIPC.1158....3M}.
The uncertainty in this number is mainly due
to the difficulty in determining stellar ages accurately at these
early times, and the uncertain duration of star formation in a cluster.
There is also a wide dispersion in the lifetimes:
some objects lose their inner disks at a very early age
($\simlt 1$\,Myr), even before they become optically revealed and can
be accurately placed in the Hertzsprung-Russell diagram, 
whereas other objects retain their accretion disks for up to 10\,Myr.

\subsection{Spitzer results: the planet-forming regions of the disk}
\label{sec:disk_lifetimes_spitzer}
Because only circumstellar dust very close to the star ($R\simlt 0.1$\,AU)
becomes hot enough to be detectable in the near-infrared, observations at
these wavelengths provide no information on the presence of circumstellar
material beyond $\sim 0.5$\,AU.  As a result, disk lifetime studies based
on near-infrared excesses always left room for the possibility that
stars without near-infrared excess had longer-lived outer disks with
enough material to form planets at radii that could not be detected
at these short wavelengths. 

\emph{IRAS} and \emph{ISO} both had the appropriate wavelength range to study
the planet-forming regions of the disk ($R\sim 0.5-20$\,AU) but lacked
the sensitivity needed to detect all but the strongest mid- and 
far-infrared excesses in low-mass stars at the distances of the nearest
star-forming regions.
During its 6-year cryogenic mission, \emph{Spitzer} provided,
for the first time, the wavelength coverage and the sensitivity needed
to detect very small amounts of dust in the planet-forming regions of
thousands of YSOs. The wealth of data made available by
\emph{Spitzer} has not only firmly established the dissipation time
scale of primordial disks, but also made it possible to address
second-order questions such as the effect of stellar mass,
multiplicity, and external environment on disk lifetimes.

\subsubsection{Stellar clusters and associations}
\label{sec:disk_lifetimes_clusters}
Just as was done for the inner disk with near-infrared studies,
the dissipation time scale of regions farther out in the disk can be
investigated by observing the fraction of stars that show mid-infrared
excesses in clusters of different ages. 
Reaching the stellar photospheres of all the targets is needed to
unambiguously establish the fraction of them that have an infrared excess
indicating the presence of a disk.  
However, while at $8.0\,\mu$m \emph{Spitzer} was able to detect the
stellar photospheres of solar-type stars at distances of up to 1\,kpc,
at longer wavelengths it was only sensitive enough to detect solar-type
photospheres within $\sim 200$\,pc.
Because there are very few stellar clusters within 200\,pc,
we focus on the results from IRAC (3.6 to 8.0\,$\mu$m) observations.

Comparing the disk fractions of stellar clusters presented by different
\emph{Spitzer} studies is not completely straightforward as the methods
used to decide which objects are members of the cluster, the disk
identification criteria, sensitivities, and  ranges of stellar masses
considered vary from study to study. 

Very  young embedded clusters (age $\leq 1$\,Myr) such as Serpens and
NGC~1333 show disk fractions of the order of 70 to 80\%
\citep{2007ApJ...669..493W, 2008ApJ...674..336G}.
This implies that pre-main sequence stars without a disk,
showing photospheric IRAC fluxes,
are seen even in extremely young star-forming regions.
Clusters in the $2-3$\,Myr range, such as IC~348 and NGC~2264,  
show IRAC disk fractions of the order of 40 to 50\%
\citep{2006AJ....131.1574L, 2009AJ....138.1116S},
whereas regions with estimated ages around $\sim 5$\,Myr, such as
Upper Scorpius and NGC~2362 exhibit disk fractions that are already
below 20\%.  By $\sim 8-10$\,Myr old, primordial disks with IRAC
excesses  become exceedingly rare as attested by their very low
incidence ($\simlt 5$\%) in regions such as the TW~Hydra, $\sigma$~Ori,
and NGC~7160 associations
\citep{2006ApJ...638..897S, 2007ApJ...662.1067H}.

The main results on disk lifetimes from IRAC studies of young stellar
clusters listed above are indistinguishable from those obtained by
pre-\emph{Spitzer} near-infrared studies.  This is not too surprising
considering that IRAC wavelengths only trace regions inward of
$\sim 5$\,AU around solar type stars.  These regions are further out
in the disk than those traced by near-infrared observations, but still exclude
the bulk of the disk material.  Thus, the question remains:
can primordial outer disks ($R>5$\,AU) survive beyond 10\,Myr?
To address this question, we must turn to the longer wavelength camera,
MIPS, on \emph{Spitzer}, and studies of star-forming regions within 200\,pc,
where $24\,\mu$m observations are sensitive enough to reach the stellar
photosphere of their targets and unambiguously identify the presence of a disk.

\subsubsection{The distributed population of pre-main sequence stars}
\label{sec:disk_lifetimes_distributed}
Firmly establishing the longevity of primordial disks was one of the
central goals of two \emph{Spitzer} Legacy Programs:
``From Molecular Cores to Planet-forming Disks''
\citep[cd2:][]{2003PASP..115..965E} and
``Formation and Evolution of Planetary Systems''
\citep[FEPS:][]{2006PASP..118.1690M}.

As part of the c2d project, \emph{Spitzer} observed over 150 WTTS
associated with the Chameleon, Lupus, Ophiuchus, and Taurus star-forming
regions, all within 200\,pc of the sun.
As one of the main goals was to establish whether the outer disk could
significantly outlive the inner disk, the c2d sample of WTTS was mostly
composed of X-ray identified and spectroscopically confirmed pre-main sequence
stars without evidence for an inner accretion disk. 

Results from the c2d project
\citep{2006ApJ...645.1283P, 2007ApJ...667..308C, 2010ApJ...724..835W}
show that $\sim 80$\% of young WTTS present $24\,\mu$m fluxes consistent
with bare stellar photospheres.  Because $24\,\mu$m observations are
sensitive to very small amounts of micron sized dust ($\ll 1\,M_{\oplus}$)
out to tens of AU from the central star, this suggests that once
accretion stops and the inner disk dissipates, the entire disk goes
away very quickly.
The fraction of c2d WTTS with $24\,\mu$m excesses is a function of
stellar age.  Approximately 50\% of the WTTS younger than $\simlt 1-2$\,Myr
do not have a disk, suggesting that a significant fraction of the entire
pre-main sequence population lose their disks by an age of $\sim 1$\,Myr.
However, none of the WTTS in the c2d sample older than
$\sim 10$\,Myr has a detectable disk.  Also, many of the WTTS disks have
very low fractional disk luminosities, ($L_{\rm disk}/L_* < 10^{-3}$),
and are thus more consistent with optically thin debris disks than with
primordial disks. 

Very similar results were obtained by the FEPS project, which observed
314 solar-type stars with a median distance of 50\,pc and ages ranging
from 3\,Myr to 3\,Gyr.
The youngest age bin in the FEPS study included 34 targets younger than
10\,Myr and a mean age of $\sim 5$\,Myr. They find that only 4 out of those
34 targets had optically thick primordial disks, whereas 5 of them had
optically thin debris disks. The second youngest bin in their study had
49 targets with estimated ages in the 10 to 30\,Myr range.
In this age bin, they identified 9 debris disk and only one primordial disk
\citep{2009ApJS..181..197C}.
The only target with a primordial disk in this age range was PDS~66,
a member of the Lower Centaurus Crux association with an estimated,
but uncertain, age of 12\,Myr \citep{2008hsf2.book..235P}.

Both the c2d and FEPS results strongly suggest that, while there is a wide
dispersion on disk lifetimes, $\sim 10$\,Myr is a firm upper limit for
the longevity of primordial circumstellar disks around solar-type stars.  
They also show that there is a significant overlap in the age distributions
of primordial and debris disks. Exactly how and when protoplanetary disks
evolve into planetary debris disks remains an open question and is
discussed in \S\ref{sec:disk_evolution_paths}.

\subsection{Dissipation timescale}
\label{sec:dissipation}
The fact that very few objects lacking near-infrared excess show
mid-infrared excess
emission implies that, once accretion stops and the inner disk clears out,
the entire disk dissipates very rapidly.  This is also supported by
(sub-)millimeter observations
\citep{1994ApJ...420..837A, 2005ApJ...631.1134A, 2007ApJ...671.1800A}
showing a very strong correlation between the detectability of a disk at
these long wavelengths and presence of an inner accretion disk.  
This implies that the vast majority of pre-main sequence stars
in any given population
are either accreting CTTS with excess emission extending all the way from
the near-infrared to the sub-millimeter or have bare stellar photospheres. 
Based on the small incidence of objects lacking an inner disk that have
evidence of an outer disk, the dissipation timescale of the entire
primordial disk once accretion stops has repeatedly been estimated 
to be $\simlt 0.5$\,Myr
\citep[e.g.,][]{1990AJ.....99.1187S,1996AJ....111.2066W, 2007ApJ...667..308C}.
This finding, that circumstellar material can survive at all radii
($R\simlt 0.1-200$\,AU) for several Myr, while later the dissipation of
the entire disk occurs  in a much shorter timescale is known as the
``two-time-scale" problem.  Disk evolution models combining viscous 
accretion with photoevaporation
\citep[e.g.,][]{2006MNRAS.369..216A, 2006MNRAS.369..229A}
have successfully been able to reproduce this behavior
(see \S\ref{sec:disk_evolution_photoevaporation}).

\subsection{Dependence on stellar mass}
\label{sec:disk_lifetimes_mass_dependence}
Young stellar clusters and associations provide an excellent opportunity
to investigate disk lifetimes as a function of stellar mass as they contain
large samples of mostly coeval stars spanning a large range of stellar masses.
In their study of the 5\,Myr old Upper Scorpius OB association,
\cite{2006ApJ...651L..49C} obtained 4.5, 8.0 and $16\,\mu$m
photometry of 204 members with masses ranging from $\sim 0.1$ to
$20\,M_{\odot}$.  They find that while $\sim 20$\% of their 127 K and M-type
targets (masses $\sim 0.1-1.2\,M_{\odot}$) are surrounded by optically
thick, primordial circumstellar disks, none of their 30 F and G-type
stars (masses $\sim 1.2-1.8\,M_{\odot}$) had any evidence for a disk
at wavelengths $\le 16\,\mu$m.

An almost identical result was obtained by \cite{2007AJ....133.2072D}
in the also 5\,Myr old cluster NGC~2362. They found a total IRAC disk
fraction of  $\sim 20$\%  (7\% of strong  excesses and 12\% of weaker excesses)
among 220 stars with estimated masses below $1.2\,M_{\odot}$ and a
complete absence of IRAC excesses among 33 stars with estimated 
masses above $1.2\,M_{\odot}$.  
MIPS observations of NGC~2362 \citep{2009ApJ...698....1C} and other 5\,Myr
old clusters such as $\lambda$ Orionis \citep{2009ApJ...707..705H}
do show $24\,\mu$m excesses for objects with masses above $1.2\,M_{\odot}$,
but they are all consistent with the optically thin emission expected from
debris disks. Overall, the \emph{Spitzer} results suggest that,
though primordial circumstellar disks can last for up to 10\,Myr around
solar and lower-mass stars, disk lifetimes are a factor of $\sim 2$
shorter around higher mass objects.  
The higher accretion rates and radiation environment are likely to be
responsible for the shorter disk lifetimes in higher mass stars
\citep{2008PhST..130a4024H, 2005AJ....129..935C, 2006A&A...459..837G}.

At the lower end of the mass distribution, the statistics are much poorer.
The disk fractions of brown dwarfs in relatively young regions (age $1-3$\,Myr)
such as Taurus, Chamaeleon~I, and IC~348 has been found to be between 40 and
50\% with statistical uncertainties as large as 10 to 20\%
\citep{2005ApJ...631L..69L, 2007A&A...465..855G, 2010A&A...515A..91M}.
For the 5\,Myr  Upper Scorpius region, the reported values
range from $37\pm 9$\% \citep{2007ApJ...660.1517S}
to $11^{+9}_{-3.3}$\% \citep{2009ApJ...705.1173R}.
Also, 3 of the 5 known brown dwarfs in the 10\,Myr old TW Hydrae
association have infrared excesses indicating the presence of a disk
\citep{2008ApJ...681.1584R}.
Taken together, these observations suggest that the dissipation time
scales of disks around substellar objects are at least as long
as those of solar-mass stars, although significantly longer disk lifetimes
cannot be ruled out based on the available data.

\subsection{Gas dispersal} 
\label{sec:disk_lifetimes_gas}
CO observations toward disks that are warm enough to prevent
freeze-out hint at a decrease in the gas-to-dust ratio with
evolutionary class \citep{2003A&A...402.1003D, 2010A&A...520A..61C}.
Unlike dust, however, there is no generic tracer of the gas mass
in young disks (see \S\ref{sec:disk_properties_gas})
and systematic studies of gas dispersal with stellar
age has been studied by observing accretion indicators.

Using the equivalent widths and velocity profiles of the H$\alpha$ line
to identify accretors, \cite{2010A&A...510A..72F} recently studied the
fractions of accreting objects in stellar clusters with ages in the
$1-50$\,Myr range. They find no accreting objects in the clusters older
than 10 Myr down to an accretion sensitivity estimated to be
$\sim 10^{-11}\,M_{\odot}\,{\rm yr}^{-1}$. 
They also find that, in most clusters,  the fraction of accreting stars
is systematically lower than the fraction of objects with \emph{Spitzer}
excesses in IRAC bands. This is expected as  $\sim 20\%$ of non-accreting
pre-main sequence stars (i.e., WTTS) have IRAC excesses
\citep{2007ApJ...667..308C, 2007ApJ...670.1337D, 2010ApJ...724..835W},
while the fraction of accreting pre-main sequence stars (i.e., CTTS)
lacking IRAC excesses is significantly  smaller, of the order of
$\simlt$ 2-5$\%$ \citep[e.g.,][]{2010ApJ...712..925C, 2010ApJ...708.1107M}.

Accretion only shows the presence of gas in the inner disk.
Consequently, our current understanding of the time scale of the gas
dispersal is in an analogous situation to that of the dust dissipation
prior to \emph{Spitzer}: although  the longevity of gaseous inner disks
is well established to be $\simlt 10$\,Myr, the possibility of
significant amounts of gas remaining for longer period of times at larger
radii still exists. However, according to the ``UV-switch" model
(Alexander et al., 2006), photoevaporation removes all
circumstellar gas very quickly ($\ll 1$\,Myr) once accretion stops
(see \S\ref{sec:disk_evolution_photoevaporation}).
\emph{Herschel} studies currently underway of sensitive gas tracers,
such as the 63.2 $\mu$m [O I] line, should be able to test this prediction.

\subsection{Environmental influences} 
\label{sec:environment}
Protoplanetary disks are detected in a range of environments from low mass,
sparsely populated molecular clouds to massive, dense stellar clusters.
In nearby, well studied, low mass star-forming regions,
there is a remarkable similarity in the average disk properties from region
to region as demonstrated from mid-infrared colors \citep{2009A&A...504..461F}
and spectra \citep{2009ApJ...703.1964F}
and also in their mass distributions and sub-millimeter colors
\citep{2005ApJ...631.1134A, 2007ApJ...671.1800A}.
This is not unexpected given that disk sizes are orders of magnitude
smaller than typical star-to-star distances except perhaps in the
densest parts of very young protoclusters and disruptive encounters
with a passing star is not a common evolutionary determinant
\citep{2001MNRAS.325..449S}.
Gravitational perturbations have a strong impact on disk evolution,
however, in close binary or multiple systems.

The median disk lifetime of 2--3\,Myr (discussed in
\S\ref{sec:disk_lifetimes}) is derived from surveys of many clusters.
There is a rather small dispersion in the disk fraction at any given
age despite the different cluster sizes and whether or not they
contain massive stars. These mid-infrared observations are unable to measure
the properties of the outer parts of the disk, however, which are more
susceptible to external influences.
Here, we discuss the influences of binary stars, massive stars,
and also metallicity on disk lifetimes.

\subsubsection{Dynamical Disruption in Binaries}
\label{sec:environment_binaries}
About half of field stars are in binary or higher
order multiple systems \citep{2010ApJS..190....1R}.
The orbital resonances in these systems have a profound influence
on disk evolution.
\cite{1994ApJ...421..651A} show that (coplanar) disks around each star
in a binary system are truncated at the outer edge,
and a circumbinary ring about both stars are truncated at the inner edge.
As a rough guide for near circular orbits with semi-major axis $a$,
the circumprimary disk is limited in size to $\sim a/2$
and the circumbinary disk's inner edge is $\sim 2a$.
Higher eccentricities lead to greater disk erosion.

There are a few resolved disk images that confirm this general pattern.
GG~Tau is perhaps one of the best studied circumbinary disks
with an inner radius of 180\,AU \citep{1999A&A...348..570G}
although it appears to be bigger than expected given the
$a=32\,{\rm AU}, e=0.34$ binary orbit \citep{2005A&A...439..585B}.
UY~Aur \citep{1998A&A...332..867D}
and CoKu~Tau~4 \citep{2008ApJ...678L..59I}
are additional examples of primordial circumbinary disks but they 
appear to be rare exceptions rather than the rule.
Outwardly truncated circumstellar disks, only a few tens of AU in
diameter, have also been detected in binary systems,
L1551-IRS5 \citep{1998Natur.395..355R},
GQ Lup \citep{2010AJ....139..626D},
HD 98800 and Hen 3-600 \citep{2010ApJ...710..462A}.

High resolution speckle imaging studies searching for a connection
between binaries and premature disk dissipation initially yielded
mixed and inconclusive results.
\cite{1993AJ....106.2005G} surveyed $\sim 70$ pre-main sequence stars
in Taurus and Ophiuchus
and concluded that the incidence of close binaries ($a < 50$\,AU)
in WTTS is enhanced with respect to that of CTTS.
This was not confirmed, however, by subsequent larger studies in Taurus
\citep{1993A&A...278..129L, 1998A&A...331..977K}.
Yet a survey of $\sim 160$ pre-main sequence stars in Ophiuchus
by \cite{2005A&A...437..611R} showed that YSOs with infrared excesses
tend to have fewer companions and are at smaller projected separations
than diskless YSOs.

Here, again, the uniformity and sensitivity of \emph{Spitzer}
studies have improved our understanding of the situation.
\cite{2009ApJ...696L..84C} combined the results from several
multiplicity surveys of pre-main sequence stars with \emph{Spitzer} 
data of four star-forming regions and showed that the distribution
of projected separations of systems with mid-infrared excesses are in fact
significantly different from that of systems without.
Binaries with projected separations less than 40\,AU are half as likely to
possess a disk than those with projected separations in the $40-400$\,AU range.
\cite{2010ApJ...709L.114D} finds a similar result with a somewhat smaller
sample.
(Sub-)millimeter fluxes are also known to be lower for similarly close
binaries indicating that the bulk of the disk has been lost
\citep{1995ApJ...439..288O, 1994ApJ...429L..29J, 1996ApJ...458..312J, 2005ApJ...631.1134A}.

Even though several factors (e.g., the incompleteness of the census of
close binaries, the use of unresolved disk indicators, and projection effects)
tend to weaken its observable signature, the effect of multiplicity on
disk lifetimes could be very strong.
The distribution of physical separations, $a$, in solar-type pre-main sequence
binaries is expected to peak around 30 AU as in field stars
\citep{1991A&A...248..485D, 2010ApJS..190....1R}.
Hence the disks around individual stars in most binary systems should
have truncation radii of the order of $(0.3-0.5)a \sim 10-15$\,AU
\citep{1977MNRAS.181..441P}.
These truncation radii are about an order of magnitude smaller than the typical
radii of disks around single stars (see \S\ref{sec:disk_properties_radius}).
For an accretion disk with $\gamma=1$
(see \S\ref{sec:disk_properties_surfacedensity}),
the viscous time scales linearly with radius and is therefore also
about an order of magnitude smaller.
This implies that the lifetimes of disks around the individual components
of most binary systems should be $\sim 10$\% of those of single stars
or about 0.3\,Myr.
These very short lifetimes for the stars in medium-separation binary
systems may explain both the presence of very young diskless stars and
the large dispersion in disk dissipation time scales.


\subsubsection{Photoevaporation by Massive Stars}
\label{sec:environment_photoevaporation}
Most stars form in large clusters with hundreds if not thousands
of stars \citep{1997ApJ...476..144M, 2003ARA&A..41...57L}.
Such large stellar groups are highly likely to contain
an O star which bathes neighboring stars and their disks in a UV radiation
field that may be several orders of magnitude above the
interstellar average. The effect is to rapidly erode the more
loosely bound outer parts of a disk.

The spectacular \emph{Hubble} images of photoevaporating disks embedded
in small ionized cocoons \citep{2000AJ....119.2919B}
make a stunning contrast to the large Taurus disks
in their more benign low radiation environment \citep{1999AJ....117.1490P}.
The pictures are somewhat misleading, however, in that the
central regions of Orion disks are not strongly affected.
Except for disks very close to the O stars where ionizing photons directly
impinge on the disk surface, the disk develops a thick photon dominated region
with a temperature $\sim 10^3$\,K.
The corresponding thermal velocity, $\sim 3$\,\kms, implies that gas is bound
to the central star for radii $\simlt 100$\,AU \citep{2004ApJ...611..360A}.
Gas pressure can cause mass loss at smaller radii but the
evaporation timescale within solar system scales, $< 50$\,AU,
is tens of Myr \citep{2007MNRAS.376.1350C}.

\begin{figure}
\centerline{\psfig{figure=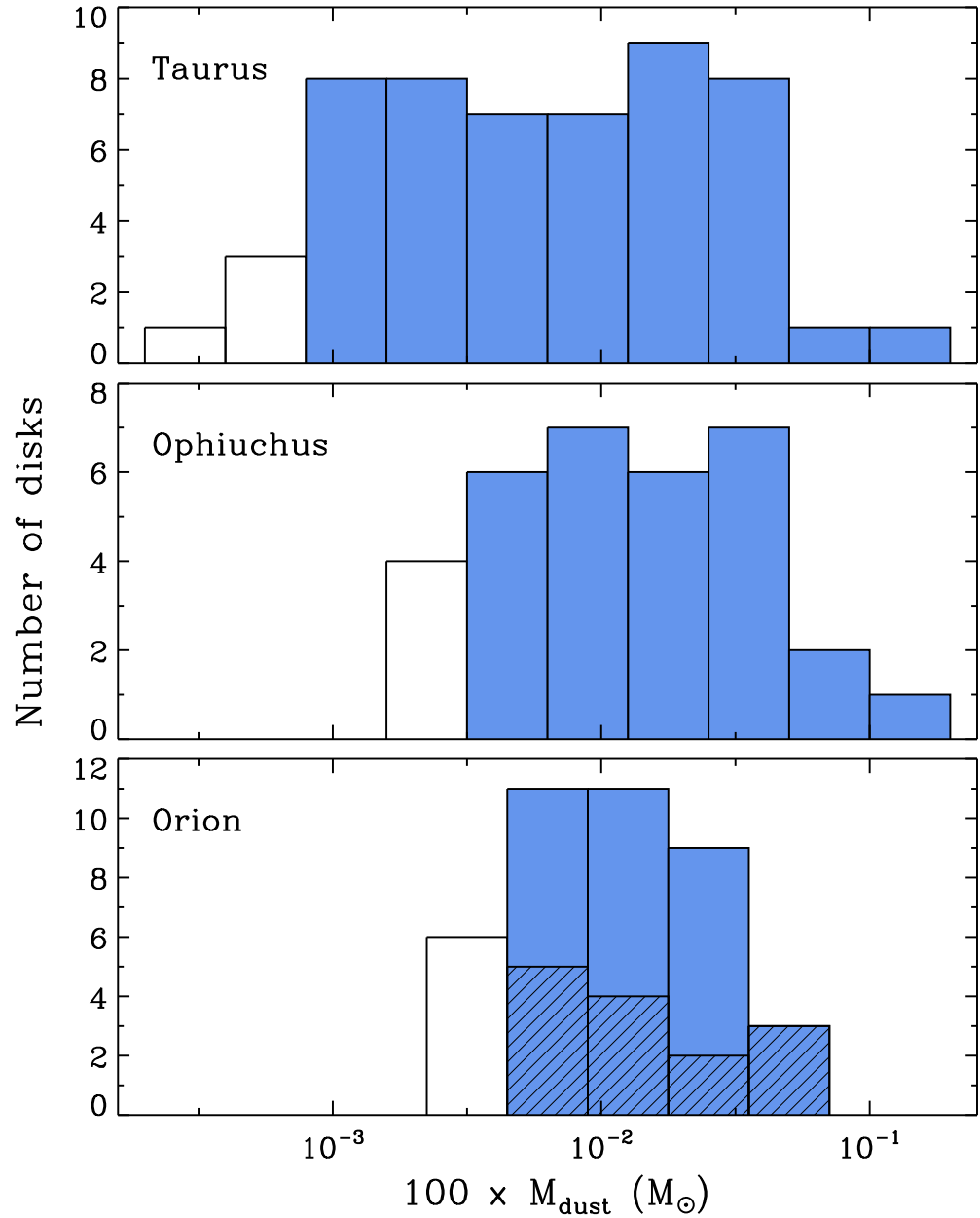,width=5.5in}}
\caption{
The distribution of protoplanetary (Class II) disk masses in
Taurus, Ophiuchus, and Orion.
The dust masses are derived from millimeter fluxes and extrapolated to
a total mass assuming a maximum grain size of 1\,mm, characteristic
temperature of 20\,K, and an interstellar gas-to-dust ratio of 100
\citep{2005ApJ...631.1134A, 2007ApJ...671.1800A, 2010ApJ...725..430M}.
The filled bars show the range where the millimeter measurements are
complete for each region. The hashed bars in the Orion histogram show
the disks with projected distances 0.3\,pc beyond
the O6 star, $\theta^1$\,Ori\,C, in the Trapezium Cluster.
}
\label{fig:diskmass_histogram}
\end{figure}

\begin{figure}
\centerline{\psfig{figure=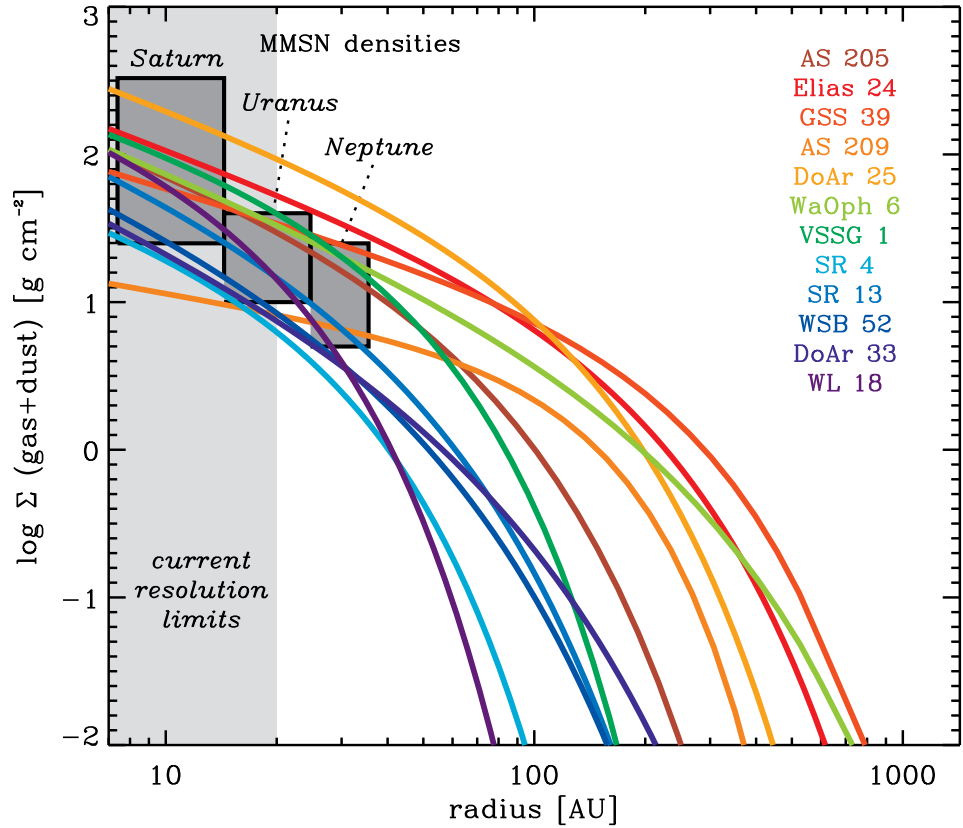,width=5.5in}}
\caption{
Radial surface density (gas+dust) profiles for Class II YSO disks in Ophiuchus
based on fitting an exponentially tapered power law profile to
$880\,\mu$m visibilities and infrared SEDs
\citep{2009ApJ...700.1502A} and \cite{2010ApJ...723.1241A}.
The dark gray rectangular regions mark the MMSN surface densities for
Saturn, Uranus, and Neptune.
}
\label{fig:sigma_sma}
\end{figure}

\begin{figure}
\centerline{\psfig{figure=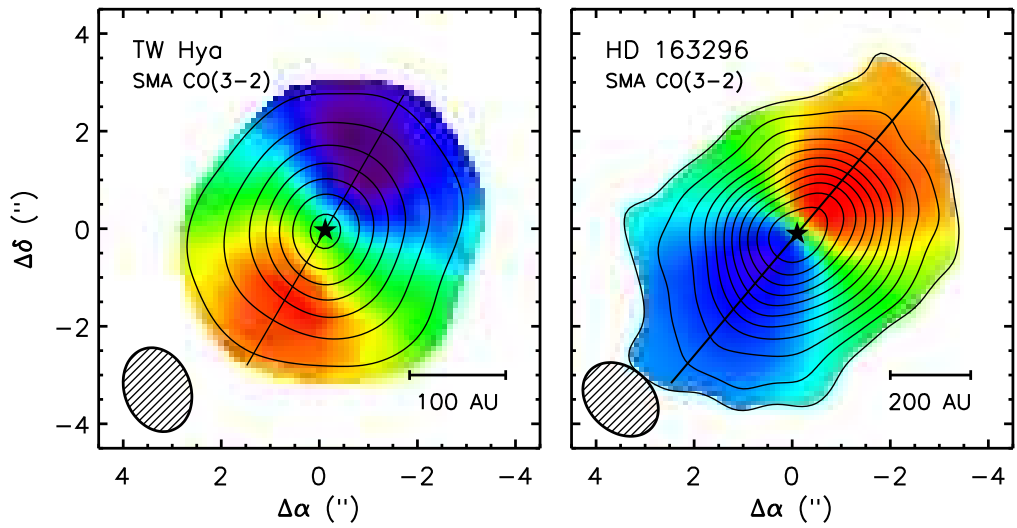,width=5.5in}}
\caption{
CO(3-2) emission from the disks around TW~Hydra (left) and HD~163296 (right)
observed with the SMA at a spectral resolution of $44\,{\rm m\,s}^{-1}$
\citep{2010arXiv1011.3826H}.
The contours show the zeroth moment (velocity-integrated intensity), whereas
the colors show the first moment (intensity-weighted velocity).
The synthesized beams are shown in the lower left corner of each panel with a
size of $1\farcs 7\times 1\farcs 3$ at position angles of $19^\circ$ and
$46^\circ$ degrees for TW~Hydra and HD~163296, respectively.
The contours start at $3\sigma$ and increase by intervals of $2\sigma$,
where the rms noise $\sigma=0.6\,{\rm Jy\,beam}^{-1}$.
}
\label{fig:momentmaps}
\end{figure}

\begin{figure}
\centerline{\psfig{figure=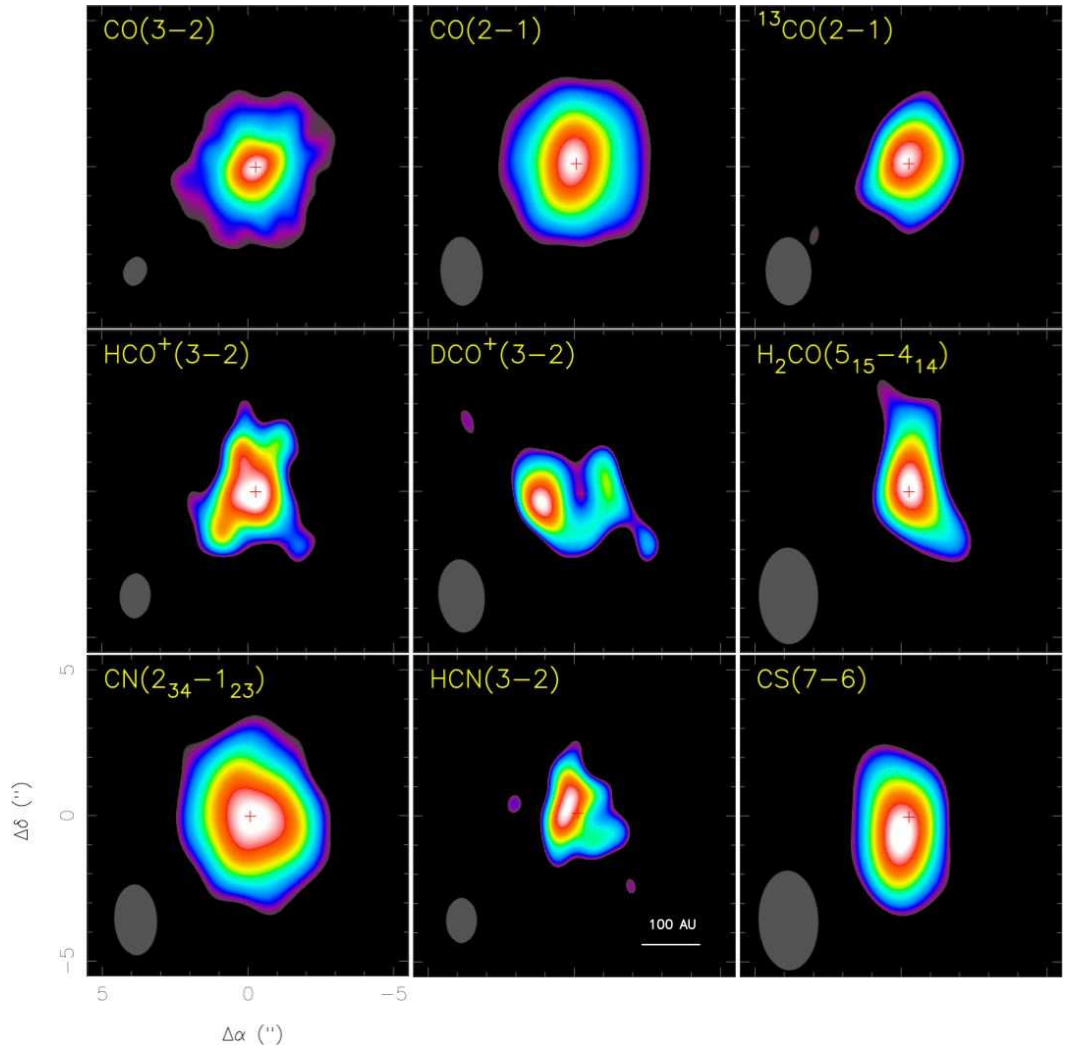,width=5.5in}}
\caption{
Sub-millimeter spectroscopy of molecular rotational lines in the
chemically rich nearby TW Hydra disk. These observations, made with
the \emph{SMA} are at a range of resolutions, shown in the lower left
corner of each panel. The nearly face-on disk generally shows centrally
peaked emission except for the DCO$^+$ line which peaks in a ring
where the temperature is colder than the inner disk and CO freezes
out of the gas phase onto grain surfaces.
(Figure courtesy of Charlie Qi.)
}
\label{fig:twhya_chemistry}
\end{figure}

\begin{figure}
\centerline{\psfig{figure=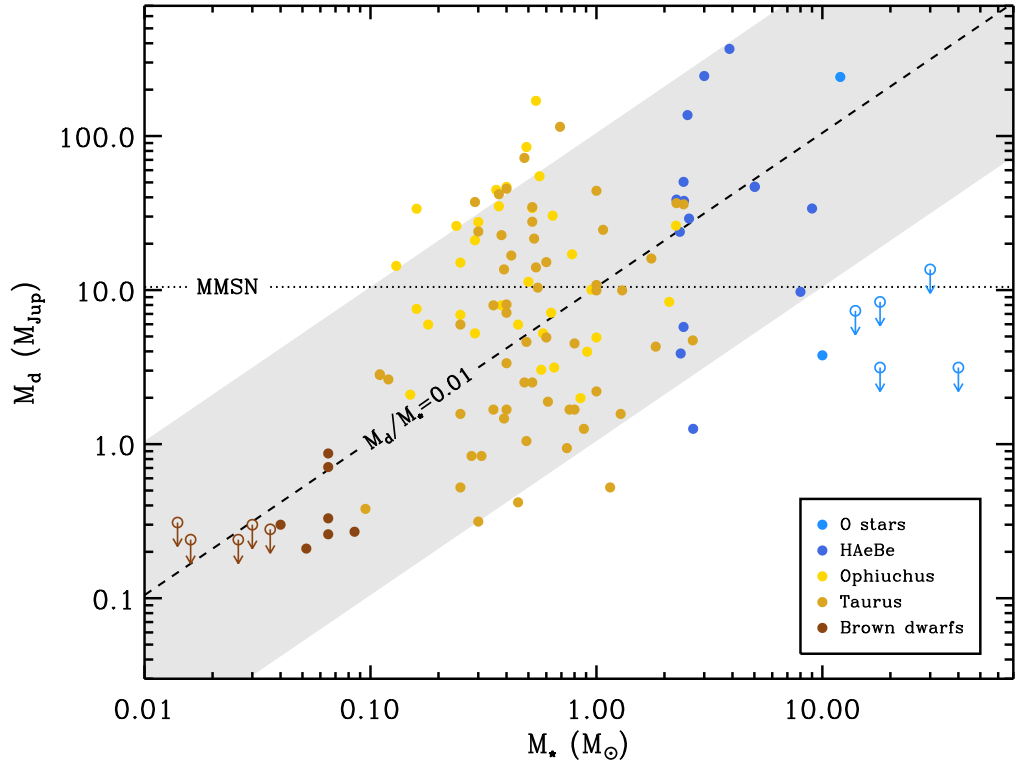,angle=90,height=4.7in}}
\caption{
The variation of protoplanetary disk mass with the mass of the central star.
Upper limits are only shown at the extremes of the stellar mass range
where no disks have been detected.
The dashed diagonal lines delineates where the mass ratio is 1\%,
and is close to the median of the detections.
Almost all the disks around stars with masses $M_*=0.04-10\,M_\odot$
lie within the grey shaded area, $\pm 1$\,dex, about the median.
The exception are O stars where no disks are detected at (sub-)millimeter
wavelengths, indicating either very short disk lifetimes or a different
star formation scenario.
}
\label{fig:mass_by_star}
\end{figure}

The comparison of disk masses in Orion with Taurus and Ophiuchus
in Figure~\ref{fig:diskmass_histogram} illustrate these points.
There is a deficit of massive disks at the center
of the Trapezium Cluster but such disks would likely have been
the largest and therefore the most susceptible to photoevaporation
by the O6 star, $\theta^1$\,Ori\,C.
The discrepancy at the upper end of the mass distribution is roughly
consistent with the smaller median size of Orion proplyds compared
to Taurus disks.
Further, the fraction of disks with at least a MMSN, $10\,M_{\rm Jup}$,
within 60\,AU is similar, $\sim 11-13$\%,
in Taurus, Ophiuchus, and Orion and, incidentally, comparable to the
detection rate of Jupiter mass extrasolar planets
\citep{2009ApJ...694L..36M}.

More massive disks are found further away from the
center of the Trapezium Cluster
\citep{2009ApJ...699L..55M, 2008ApJ...683..304E},
and the disk mass distribution of the full Orion region more closely
resembles those in Taurus and Ophiuchus
(see Figure~\ref{fig:diskmass_histogram}).
The large difference in mass between the most massive disks in and out of
the cluster center compared to the difference in stellar density
rules out stellar encounters as the cause \citep{2006ApJ...642.1140O}.

\emph{VLA} and \emph{SMA} observations do not currently have
sufficient sensitivity to study more distant regions.
However, \emph{Hubble} and \emph{Spitzer} observations find
morphological features similar to the Orion proplyds in other
massive star-forming regions
\citep{2003ApJ...587L.105S, 2006ApJ...650L..83B}.
\cite{2008ApJ...686.1195H} find that the $\gamma$ Velorum cluster
has a low fraction of sources with infrared excesses for its
age and consider disk photoevaporation as a possibility
but also note the large uncertainty in the cluster age.
The disk fraction decreases by about a factor of 2 within
the central 0.5\,pc of the Rosette nebula \citep{2007ApJ...660.1532B}
and the central 1\,pc of S~Monoceros \citep{2009AJ....138.1116S}
but the statistics are limited by small numbers.
External evaporation can only remove the inner disk on such
short timescales with very strong radiation fields that directly
ionize the disk and boil it off at $\sim 10^4$\,K
\citep{2004ApJ...611..360A, 2007MNRAS.376.1350C}.
It is hard to reconcile this with a parsec-scale sphere of
influence of a massive star unless the stellar orbits are highly
eccentric \citep{1999ApJ...515..669S}.

\cite{2008ApJ...688..408B} show that the Spitzer $24\,\mu$m dusty tails
in these regions are not detected in Pa$\alpha$ and are effectively
gas-free. They suggest that photoevaporation removes the gas very
quickly but that a dusty reservoir is replenished by the collisions
of large bodies left behind in the disk.
\cite{2005ApJ...623L.149T} consider the rapid enhancement of the dust-to-gas
ratio in photoevaporating disks as a potential trigger for planetesimal
formation. External evaporation has also been considered as a possible
explanation of the steep drop in the surface density of Kuiper Belt
Objects beyond 50\,AU \citep{1998AJ....115.2125J}.

\subsubsection{The effect of metallicity}
\label{sec:environment_metallicity}
The effect of the metallicity, or the initial dust-to-gas content,
on the evolution of protoplanetary disks has not been well studied,
at least at the early stages. It is known that the metallicity of the
host star is strongly correlated with the presence of hot Jupiters
\citep{2005ApJS..159..141V}.
This may be due to an enhanced rate of planetesimal formation
\citep{2009ApJ...704L..75J}
but longer disk lifetimes may also play a role.
In a small sample, \cite{2009A&A...501..973D} do not find any significant
difference in the metallicities of Taurus and Orion WTTS or CTTS.
However, \cite{2009ApJ...705...54Y} showed the the inner disk
fraction, as measured by near-infrared excesses, is significantly smaller
in low metallicity clusters at the edge of the Galaxy.
\cite{2010ApJ...723L.113Y} extend this work to other distant, low metallicity
clusters of different ages and find a median disk lifetime, $< 1$\,Myr.
They suggest a strong dependence of the lifetime on metallicity,
which is in qualitative agreement with photoevaporative models of
disk evolution \citep{2010MNRAS.402.2735E}.

\subsection{Section summary}   
\begin{itemize}     
 \item Disk lifetimes have a median between 2 and 3\,Myr but
       vary from less than 1\,Myr to a maximum of 10\,Myr.
       Multi-wavelength observations show that the dissipation
       timescale is very similar at all radii.
 \item Circumstellar disks dissipate faster around high-mass stars than
       around solar mass stars. Very low mass-stars and brown dwarfs have
       disk lifetimes at least as long as, and likely longer than,
       those of solar mass stars. 
 \item Disk lifetimes are significantly shorter around medium separation
       binaries ($a\sim 5-100$\,AU).
 \item Photoevaporation from massive stars erodes the outer parts of
       disks but generally leaves their interiors ($\simlt 50$\,AU) intact.
       The lifetimes and masses are only significantly reduced for
       disks that lie within a few tenths of a parsec from an O star.
 \item Disks in low metallicity environments at the edge of the Galaxy
       have median lifetimes $< 1$\,Myr, which are substantially shorter
       than in the Solar neighborhood.
\end{itemize}

\section{DISK EVOLUTION}
\label{sec:disk_evolution}
Understanding the physical processes that drive the evolution of
primordial circumstellar disks, as they evolve from optically thick
to optically thin, is crucial for our understanding of planet formation.
Disks evolve through various processes, including viscous accretion,
dust settling and coagulation, dynamical interactions with (sub-)stellar
companions and forming planets, and photo-evaporation by ultraviolet
and X-ray radiation.
In this section we summarize the  models and observational constraints
for the different processes that control the evolution of primordial
circumstellar disks.

\subsection{Viscous transport}
\label{sec:disk_evolution_viscous}
To first order, the evolution of primordial disks is driven by viscous
transport.
The accretion from the inner disk onto the star is relatively well understood
and well constrained observationally. The large velocity widths,
intensities, and profiles of emission lines such as H$\alpha$,
Br$\gamma$ and Ca II can be successfully reproduced by
magnetospheric accretion models \citep{1998ApJ...492..743M}.
The physical mechanisms that drive the radial transport across
the disk are reviewed by \cite{2010arXiv1011.1496A}.

Viscous evolution models are broadly consistent with the observational constraints
for disk masses and sizes, and the decrease in accretion rate over time
\citep{1998ApJ...495..385H, 2005A&A...442..703H}.
However, pure viscous evolution models also predict
a smooth, power-law evolution of the disk properties.
This secular disk evolution is inconsistent with the very rapid disk
dissipation that usually occurs after a
much longer disk lifetime (i.e. the ``two-time-scale" problem). 
Viscous evolution models also fail to explain the variety of SEDs
observed in the transition objects discussed in \S\ref{sec:transition_disks}.
These important limitations of the viscous evolution models show
that they  are in fact just a first-order approximation of a much more
complex evolution involving several other important physical processes.

\subsection{Photoevaporation by radiation from the central star}
\label{sec:disk_evolution_photoevaporation}
Together with viscous accretion, photoevaporation is one of the
main mechanisms through which primordial circumstellar disks
are believed to lose mass and eventually dissipate.  Photoevaporation
can be driven by energetic photons in the
far-ultraviolet (FUV: 6\,eV $< h\nu <$ 13.6\,eV), 
extreme-ultraviolet (EUV: 13.6\,eV $< h\nu <$ 0.1\,keV)
and X-ray ($h\nu > 0.1$\,keV) energy range.
Photons in each energy range affect the disks in different ways,
and the relative importance of FUV, EUV, and X-ray photoevaporation
is still not well understood. Photoevaporating photons can originate
both from nearby massive stars and from the central star itself.
The former scenario was discussed in \S\ref{sec:environment_photoevaporation}
and here we focus on the latter.

Early photoevaporation models focused on the effect of ionizing EUV radiation 
\citep[e.g.,][]{1994ApJ...428..654H}
on circumstellar gas around early-type stars. 
Later disk evolution models, known as ``UV-switch'' models,
combine viscous evolution with photoevaporation by EUV photons
\citep{2001MNRAS.328..485C, 2006MNRAS.369..216A, 2006MNRAS.369..229A}
to tackle the ``two-time-scale" problem of T Tauri evolution
(i.e., the sudden dispersion of the entire disk after much longer
disk lifetimes, see \S\ref{sec:dissipation}).
According to these models, extreme ultraviolet (EUV) photons originating at
the stellar chromospheres of low-mass stars ionize and heat the circumstellar
hydrogen to $\sim 10^4$\,K. Beyond a critical radius,
$\sim 10$\,AU for solar mass stars,
the thermal velocity of the ionized hydrogen exceeds its escape velocity
and the material is lost in the form of a wind.  

At early stages in the evolution of the disk, the accretion rate dominates
over the evaporation rate and the disk undergoes standard viscous evolution:
material from the inner disk is accreted onto the star, while the outer disk
behaves as a reservoir that resupplies the inner disk, spreading as angular
momentum is transported outwards. Later on, as the accretion rate drops to the
photoevaporation rate, $\sim 10^{-10}-10^{-9}\,M_{\odot}\,{\rm yr}^{-1}$
in the models, the outer disk is no longer able to resupply the inner
disk with material. At this point, the inner disk drains on a viscous
timescale ($\simlt 10^5$\,yr) and an inner hole of a few AU in radius
is formed in the disk.  The inner disk edge is now directly exposed to the
EUV radiation and the disk rapidly photoevaporates from the inside out. 
Thus, the UV-switch model naturally accounts for the lifetimes and
dissipation timescales of disks as well as for SEDs of some
pre-main sequence stars suggesting the presence of large inner holes. 

Recent photoevaporation models include X-ray
\citep[e.g.,]{2010MNRAS.401.1415O}
and/or FUV irradiation 
\citep[e.g.,]{2009ApJ...705.1237G, 2009ApJ...690.1539G}
in addition to the EUV photons.  
Because X-rays and FUV photons are able to penetrate much larger columns
of neutral gas than EUV photons, they are able to heat gas that is
located both deeper in the disk and at larger radii.
Thus, while EUV induced photoevaporation is restricted to the inner
few AU of the disk, X-rays and FUV photons can operate at tens of AU
from the star.

EUV+X ray and FUV+X-ray models show a similar qualitative behavior
to photoevaporation by EUV alone, but also several important differences
\citep{2010arXiv1011.1496A}.
Most notably, they predict photoevaporation rates of the order of
$10^{-8}\,M_\odot\,{\rm yr}^{-1}$, which are two orders of magnitude 
greater than pure EUV photoevaporation models.
Consequently an inner hole forms early in a disk's accretion history.
As a result, EUV+X ray and FUV+X-ray models predict a significant
population of pre-main sequence stars with relatively massive
($\simgt 10\,M_{\rm Jup}$) disks with large inner holes and no,
or very little, accretion.  However, observations have shown that
disks around WTTS tend to have much lower masses, $\simlt 1-2 M\_{\rm Jup}$
\citep{2005ApJ...631.1134A, 2007ApJ...671.1800A, 2008ApJ...686L.115C, 2010ApJ...712..925C}.
This suggests that, in reality, the photoevaporation rates and disk
masses are lower than those predicted by
\cite{2009ApJ...705.1237G} and \cite{2010MNRAS.401.1415O}
when the inner hole is initially formed and that the
quantitative predictions of current photoevaporation models should
be taken with caution.

\subsection{Grain growth and dust settling}
\label{sec:disk_evolution_graingrowth}
Even though solid particles only represent 1\% of the initial 
mass of the disk, understanding their evolution is of utmost interest
for disk evolution and planet formation studies.  Solids not only dominate
the opacity of the disk, but also  provide the raw material from which
the terrestrial planets and the cores of the giant planets (in the core
accretion model) are made. Although viscous accretion and the photoevaporation
processes discussed above drive the evolution of the gas,
other processes operate on the solid particles, most importantly,
grain growth and dust settling. 

Grain growth and dust settling are intimately interconnected processes. 
Gas motions differ slightly from Keplerian motions due to pressure.
Small ($r\sim 0.1\,\mu$m) grains have a large surface-to-mass ratio
and are swept along with the gas.
As grains collide and stick together, their surface-to.mass ratio
decreases and their motions decouple from the gas. They therefore
suffer a strong drag force and settle toward the midplane.
This increases the density of dust in the interior of the disk, 
which accelerates grain growth, and results in even larger grains settling
deeper into the disk. If this process were to continue unimpeded,
the end result would be a perfectly stratified disk with only small
grains in the disk surface and large bodies in the  midplane.
However, because circumstellar disks are known to be turbulent,
some degree of vertical stirring and mixing of grains is expected
\citep{2005A&A...434..971D}.

\subsubsection{Models}
\label{sec:disk_evolution_graingrowth_models}
Grain growth represents the baby steps toward planet formation.
Over 13 orders of magnitude in linear size separate sub-micron
particles from terrestrial planets, however, and many poorly understood
processes operate along the way.  Idealized dust coagulation models,
ignoring fragmentation and radial drift, predict extremely efficient
grain growth.
\cite{2005A&A...434..971D} investigated the dust coagulation process
in circumstellar disks coupled to the settling and turbulent mixing of
grains.  They included 3 relatively well understood dust coagulation
mechanisms (Brownian motion, differential settling and turbulence)
and conclude that these processes are efficient enough to remove all
small grains ($r < 100\,\mu$m) within $10^4$\,yr.
This is clearly inconsistent with the wealth of observational evidence
showing the presence of micron sized grains throughout the duration of
the primordial disk phase discussed in the next section.
The conclusion is that small grains must be replenished
and that the persistence of small grains depends on a complex balance
between dust coagulation and fragmentation \citep{2008A&A...487..205D}.
	
Recent, more realistic models including fragmentation and radial drift confirm
the necessity of grain fragmentation to explain the ubiquity of small grains
in disks \citep{2008A&A...480..859B, 2011A&A...525A..11B}.
These same models also
confirm the severity of the problem known as the ``meter-size barrier",
a physical scale at which solids are expected to suffer both destructive
collisions and removal through rapid inward migration
\citep{1977MNRAS.180...57W}.
Even though several possible solutions have been proposed, including the
formation of planetesimals in long-lived vortices
\citep[e.g.,][]{2010MNRAS.tmp.1234H},
or via the gravitational instability of millimeter-sized chondrules in a gas
deficient subdisk \citep[e.g.,][]{2002ApJ...580..494Y}, overcoming this
barrier remains one of the biggest challenges
for planet formation theories \citep{2010AREPS..38..493C}.

\subsubsection{Evidence for grain growth from sub-micron to microns}
\label{sec:disk_evolution_graingrowth_microns}

Pristine dust in the interstellar medium is composed primarily
of amorphous silicates, generally olivine and pyroxene
with characteristic features at 9.7 and 18.5\,$\mu$m 
from Si--O stretching and O--Si--O bending modes, respectively.
The shape of these features is a sensitive diagnostic of the size
of small grains, generally in the $r\sim 0.1-5\,\mu$m range.
Although the emission features of the smallest grains are strong and narrow,
those of larger particles are weaker and broader.
Thermal annealing can modify the lattice structure of magnesium-rich
olivine and pyroxene and turn them into their crystallized forms,
known as enstatite and forsterite \citep{2010ARA&A..48...21H}.
These crystalline silicates have multi-peak features at
slightly longer wavelengths than their amorphous counterparts.  

Recent surveys with the IRS on \emph{Spitzer} have allowed
the study of silicates in hundreds of circumstellar disks in nearby
star-forming regions
\citep{2006ApJ...639..275K, 2006ApJS..165..568F, 2009ApJ...703.1964F,
2010ApJ...714..778O, 2010ApJS..188...75M}.
Because these observations trace the warm optically thin disk ``atmosphere'',
and the larger particles are expected to settle toward the disk interior,
they provide information on the smallest
population of grains present in the disk.
Nevertheless, the observed silicate features in most disks are
consistent with the presence of micron-sized particles and the
absence of sub-micron dust grains. This implies either that grain
growth is more efficient than fragmentation at these scales or
that sub-micron grains are efficiently removed from the upper layers
of the disk by stellar winds or radiation pressure
\citep{2009A&A...507..327O}.

These \emph{Spitzer} surveys have also shown that the signatures of
grain growth and crystallization are seen at very early stages of
disk evolution, even before the envelope has dissipated
\citep{2009ApJ...703.1964F, 2010ApJS..188...75M}.
There seems to be little connection between the age or evolutionary
stage of a primordial disk and the dust properties revealed by the
silicate features.  Even though many studies have searched for a
correlation between the large scale properties of the disk
(e.g., mass, accretion rates) and grain characteristics
(sizes, degree of crystallization), no conclusive evidence has yet
been found
\citep[e.g.,][]{2006ApJ...639..275K, 2009ApJ...703.1964F, 2010ApJ...714..778O}.
This lack of correlation between age and dust properties suggests that the
characteristics of the dust population depend on a balance between
grain growth and destruction and between crystallization
(via thermal annealing) and amorphization
\citep[e.g., via X-ray irradiation,][]{2009A&A...508..247G}.
This balance seems to persist throughout the duration of the primordial
disk stage, at least in the surface layers of the disk.

\subsubsection{Evidence for grain growth from microns to millimeters}
\label{sec:disk_evolution_graingrowth_mm}
A second, independent line of evidence for grain growth is found in
the slope, $\alpha_{\rm mm}$, of the SED at sub-millimeter wavelengths,
$F_\nu\propto\nu^{\alpha_{\rm mm}}$.
The slope between $\lambda\approx 0.5-1$\,mm is significantly shallower
in protoplanetary disks, $\alpha_{\rm mm}\approx 2-3$
\citep{1991ApJ...381..250B, 1994MNRAS.267..361M, 2005ApJ...631.1134A, 2007ApJ...671.1800A}
than in the diffuse interstellar medium,
$\alpha_{\rm mm}\sim 4$ \citep{1996A&A...312..256B}.
In the Rayleigh-Jeans limit, the flux,
$F_\nu\propto B_\nu (1-e^{-\tau_\nu})\rightarrow \nu^2$ for
optically thick emission and $\rightarrow \kappa_\nu\nu^2$
for optically thin emission.
The SED slope is therefore bound between $\alpha_{\rm mm}=2$ and $2+\beta$
(see equation~\ref{eq:kappa}).
Unresolved photometric observations include some optically thick
emission from the inner disk so the connection between the observed
disk average $\alpha_{\rm mm}$ and the grain opacity index, $\beta$,
requires modeling the surface density profile \citep{1990AJ.....99..924B}.
For a simple power law thin disk model, \cite{2005ApJ...631.1134A}
show that $\alpha_{\rm mm}-\beta$ lies between 1.1 and 1.8
for disk masses $10^{-2}-10^{-5}\,M_\odot$ respectively,
and is approximately independent of $\alpha_{\rm mm}$.
\cite{1991ApJ...381..250B} and \cite{1994MNRAS.267..361M}
make similar corrections and all reach the same conclusion that the
grain opacity index is significantly lower in disks,
$\beta_{\rm disk}\approx 0.5-1$,
than in the interstellar medium, $\beta_{\rm ISM}\approx 1.7$
\citep{2001ApJ...554..778L}.

The decrease in the dust opacity index
is best explained by the presence of substantially
larger dust grains in disks relative to the interstellar medium.
For a power law distribution of grain sizes, $n(a)\propto a^{-p}$,
\cite{2001ApJ...553..321D} and \cite{2004A&A...416..179N} show that
$\beta\approx 1.7$ if the maximum grain size $a_{\rm max}<\,30\mu$m
but then decreases to $\beta\simlt 1$ for $a_{\rm max}\simgt 0.5$\,mm
and $p=2.5-3.5$.
\cite{2006ApJ...636.1114D} provides a useful relation,
$\beta\approx (p-3)\beta_{\rm ISM}$
if $a_{\rm max}$ is more than 3 times the observing wavelength,
largely independent of grain composition.
The implication of shallow disk SED slopes at sub-millimeter wavelengths
relative to the diffuse interstellar medium is that dust grains have grown
by at least 3 orders of magnitude, from microns to millimeters.

Some grain growth appears to happen in dense molecular cores
before disks form.
Growth to micron sizes is inferred from ``coreshine'',
or scattering, of mid-infrared light \citep{2010Sci...329.1622P}.
\cite{2000ApJS..131..249S} measured the spectral slope for 21 cores,
some pre-stellar, others in very early stages of low mass star formation
and find an average spectral slope between 450 and $850\,\mu$m,
$\alpha_{\rm mm}=2.8$, which is intermediate between the diffuse
interstellar medium and protoplanetary disks.
Multi-wavelength observations of Class 0 cores by
\cite{2007ApJ...659..479J} and \cite{2009ApJ...696..841K} 
show opacity indices, $\beta\approx 1$, also shallower than
the diffuse interstellar medium but not quite as steep as in
protoplanetary disks.

There is also evidence that grain growth continues throughout the
disk lifetime.
\cite{2005ApJ...631.1134A} find a significant difference
in the SED slope between Taurus Class I (median $\alpha_{\rm mm}=2.5$)
and Class II YSO disks (median $\alpha_{\rm mm}=1.8$).
They also show that the sub-millimeter and infrared SED slopes are
correlated with a best fit linear relationship,
$\alpha_{\rm mm}=2.09-0.40\alpha_{\rm IR}$.
Confirmation of this single-dish result, which may be biased by cloud
contamination in the early phases, requires interferometry.

These sub-millimeter data only place a lower bound on the size of the
most massive grains. Longer wavelength observations constrain
the presence of larger grains but the flux decreases
(as the surface area of the dust decreases for a given mass)
and the free-free emission from an ionized stellar wind often
dominates for $\lambda\simgt 1$\,cm \citep{2004A&A...416..179N}.
However, the correction for an opaque inner region is smaller.
Surveys of Taurus disks have been carried out at 7\,mm by
\cite{2006A&A...446..211R} and at 3\,mm by \cite{2010A&A...512A..15R}
who both find that the shallow SED slope generally extends to their
observing wavelength and infer average opacity indices,
$\langle\beta\rangle=1,0.6$ respectively.
Similar results are found in other regions
\citep{2010A&A...521A..66R, 2011A&A...525A..81R, 2007A&A...462..211L}.
The implication is that centimeter sized particles are commonplace.
The longest wavelength detection of dust in a protoplanetary disk
to date was carried out by \cite{2005ApJ...626L.109W}.
By resolving the nearby, bright TW Hydra disk and showing that its
flux is constant with time, they convincingly show that the
3.5\,cm emission is from thermal dust emission and not accretion
shocks or a stellar wind.
The \cite{2006ApJ...636.1114D} formulation would therefore indicate
the presence of $\sim 10$\,cm sized bodies in the disk out at least
to the 15\,AU beam.
A similar result has been found for WW Cha \citep{2009A&A...495..869L}.

Collisional grain growth \citep{2008ARA&A..46...21B}
is expected to have a strong radial dependence due to the
decreasing density and rotational velocity with increasing radius.
\cite{2010ApJ...714.1746I} resolve the RY Tau and DG Tau disks at
$\lambda=1.3$ and 2.8\,mm and model the data with a radially
dependent dust opacity. However, to within their precision, 
$\Delta\beta=0.7$, they do not see a significant gradient.
We will soon see detailed images of $\beta$ in disks and learn
much more about the radial variation of grain growth via
multi-wavelength observations with \emph{ALMA}.

\subsubsection{Evidence for dust settling}  
\label{sec:disk_evolution_dust_settling}
Protoplanetary disks are flared (see \S\ref{sec:disk_properties_scaleheight})
and hence intercept and reprocess more stellar radiation than
physically thin disks.
Nevertheless, most T tauri stars exhibit less mid-infrared
emission than expected for a disk in hydrostatic equilibrium. 
This can be understood in terms of dust settling which reduces
the scale height and flaring angle of the disk \citep{2004A&A...421.1075D}.
Mid-infrared slopes can therefore be used as a diagnostic of dust settling.
\cite{2006ApJ...638..314D} explored parametric models with two populations
of grains: micron sized grains with a reduced dust-to-gas mass ratio in
the  surface layers of the disk, and  mm-sized grains with an increased
dust-to-gas mass ratio in the disk midplane. 
In order to quantify the degree of dust settling,
they introduced the parameter $\epsilon$, the ratio of the dust-to-gas
mass ratio in the surface layer of the disk to the standard dust-to-gas
mass ratio of the interstellar medium (1:100).
In this context, $\epsilon=1$ implies that no dust settling has occurred,
and $\epsilon$ decreases as dust settling increases.
\cite{2006ApJ...638..314D} find that the median mid-infrared slopes of
CTTS SEDs imply that $\epsilon\le 0.1$ and possibly as low as $10^{-3}$.

\cite{2006ApJS..165..568F} applied these parametric models to
reproduce the spectral slopes of a sample of over 80 Taurus
T Tauri stars observed with the \emph{Spitzer} Infrared Spectrograph.
They conclude that most objects are consistent with dust depletion
factors in the surface layers of the disk of order of 100 to 1000
(i.e., corresponding to  $\epsilon = 10^{-2}-10^{-3}$).
A similar result is found by \cite{2010ApJS..188...75M} in the Ophiuchus
molecular cloud.  They also  find evidence for significant dust settling
in  young (age $\sim$  0.3 Myr) objects embedded in the cloud core,
suggesting that this process sets in early in the evolution of
the disk \citep[see also][]{2009ApJ...703.1964F}.

\subsection{Typical evolution and diversity of evolutionary paths}
\label{sec:disk_evolution_paths}
Although we are far from fully understanding the complex evolution of
protoplanetary disks, a coherent picture is starting to
emerge from the many models and observational constraints discussed
above. Although it is clear that not all disks follow the same evolutionary
path, many observational trends suggest that most objects do follow
a common sequence of events.  
In what follows we summarize our current understanding of the ``typical''
evolution of a protoplanetary  disk.

\subsubsection{The evolution of a typical disk}
\label{sec:disk_evolution_typical}
\begin{figure}
\centerline{\psfig{figure=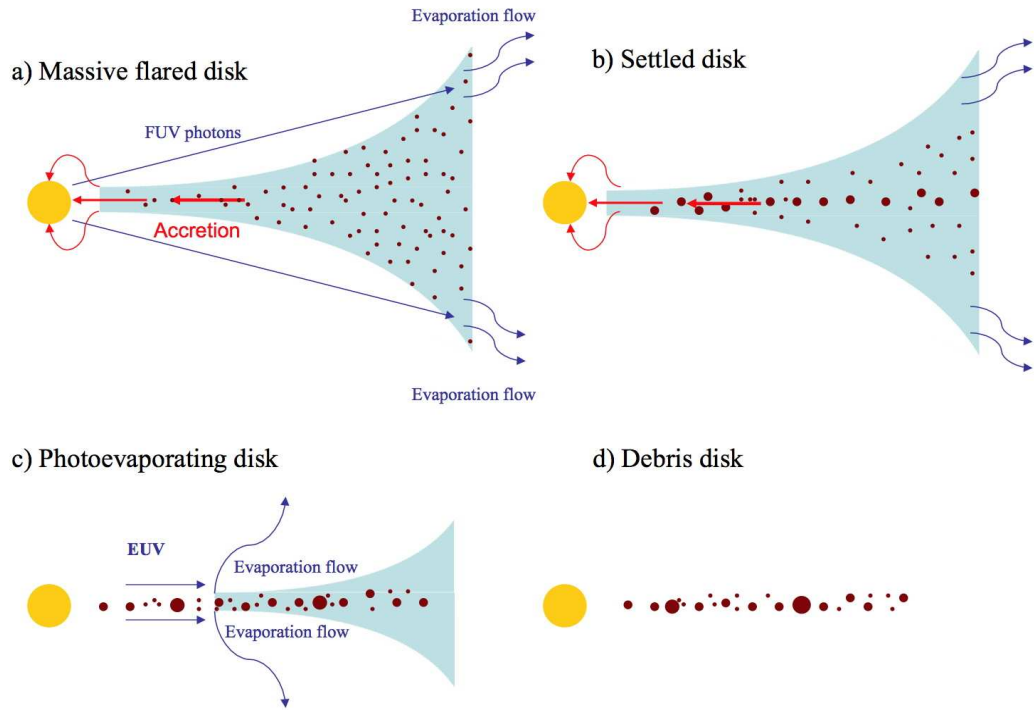,width=5.5in}}
\caption{
The evolution of a typical disk. The gas distribution is shown in
blue and the dust in brown.
(a) Early in its evolution, the disk loses mass through accretion 
onto the star and FUV photoevaporation of the outer disk.
(b) At the same time, grains grow into larger bodies that settle
to the mid-plane of the disk.
(c) As the disk mass and accretion rate decrease, EUV-induced
photoevaporation becomes important, the outer disk is no longer 
able to resupply the inner disk with material, and the inner disk
drains on a viscous timescale ($\sim 10^5$\,yr).
An inner hole is formed, accretion onto the star ceases,
and the disk quickly dissipates from the inside out.
(d) Once the remaining gas photoevaporates, the small grains are
removed by radiation pressure and Poynting-Robertson drag.
Only large grains, planetesimals, and/or planets are left 
This debris disk is very low mass and is not always detectable.
}
\label{fig:evolution}
\end{figure}

Early in its evolution, the disk loses mass through accretion onto the star
and FUV photoevaporation of the outer disk \citep{2009ApJ...705.1237G}.
The FUV photoevaporation is likely to truncate the outer edge of the disk,
limiting  its viscous expansion to a finite size of several hundreds of
AU in diameter (Figure~\ref{fig:evolution}a).
During this ``mass depletion"  stage,  which can last several Myr,
an object would be classified as a CTTS based on
the presence of accretion indicators. Accretion may be
variable on short timescales, but show a declining long-term trend.

\begin{figure}
\centerline{\psfig{figure=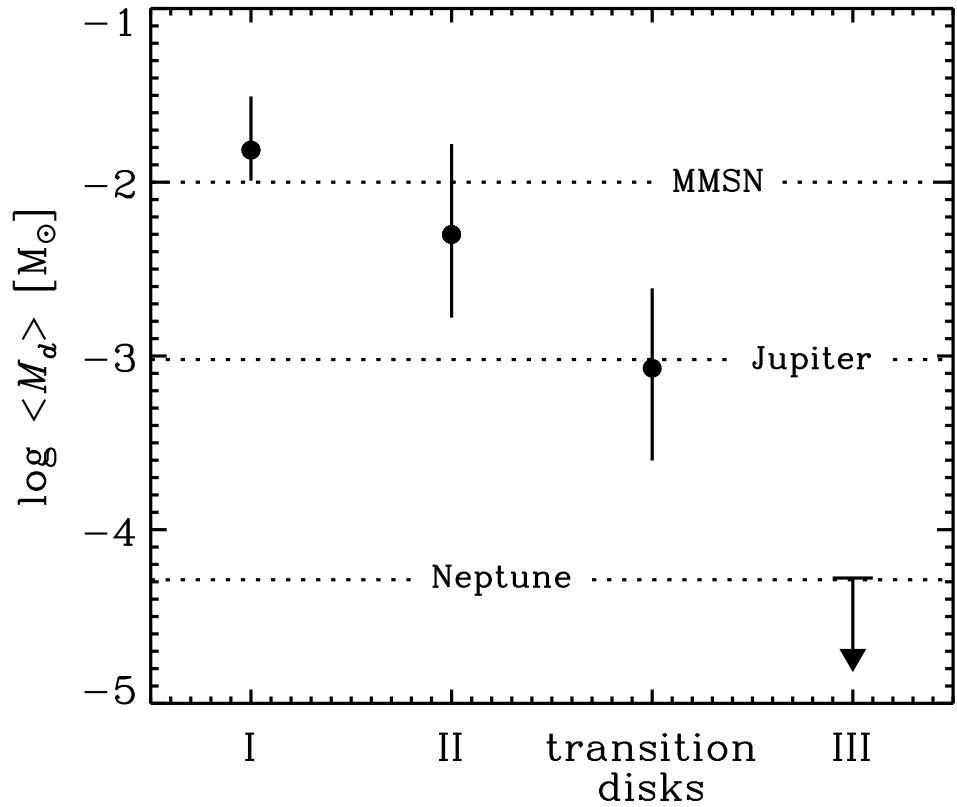,width=5in}}
\caption{
Evolution of the disk masses across the empirical sequence defined
by the slope of the infrared SED. The circles show the median
disk mass, as measured at sub-millimeter wavelengths for a
sample of 300 YSOs in Taurus and Ophiuchus. Error bars show
the distribution quartiles. Transition disks are defined here
as those YSOs that lack infrared excesses for wavelengths
less than $25\,\mu$m, but have detectable sub-millimeter emission.
With this definition, no Class III YSO was detected and the
stringent limit to their median mass comes from stacking
the non-detections together \citep{2007ApJ...671.1800A}.
}
\label{fig:mass_by_class}
\end{figure}

At the same time, grains grow into larger bodies that settle onto the
mid-plane of the disk where they can grow into rocks, planetesimals
and beyond.  Accordingly, the scale height of the dust decreases and
the initially-flared dusty disk becomes flatter (Figure~\ref{fig:evolution}b).
This steepens the slope of the mid and far-infrared SED as a smaller fraction
of the stellar radiation is intercepted by circumstellar dust 
\citep{2005A&A...434..971D}.
The near-infrared fluxes remain mostly unchanged because the inner disk
stays optically thick and extends inward to the dust sublimation
temperature. The most noticeable SED change during this stage is seen
in the decline of the (sub-)millimeter flux, which traces the decrease
in the mass of millimeter and smaller sized particles
\citep[][Figure~\ref{fig:mass_by_class}]{2005ApJ...631.1134A, 2007ApJ...671.1800A}.

As disk mass and accretion rate decrease, energetic photons from the stellar
chromosphere are able to penetrate the inner disk and photoevaporation
becomes important.  When the accretion rate drops to the photoevaporation
rate, the outer disk is no longer able to resupply the inner disk with
material \citep{2006MNRAS.369..216A, 2010MNRAS.401.1415O}.
At this point, the inner disk drains on a viscous timescale
($\simlt 10^5$\,yr) and an inner hole of a few AU in radius is
formed in the disk (Figure~\ref{fig:evolution}c).
Once this inner hole has formed, the energetic photons impact the inner
edge of the disk unimpeded, and the photoevaporation rate increases
further, preventing any material from the outer disk from flowing into
the inner hole. This halts accretion and results in the rapid transition
between the CTTS and the WTTS stage. 

The formation of the inner hole marks the end of the slow  ``mass depletion"
phase and the beginning of the of the rapid ``disk dissipation" stage.
By the time the inner hole is formed, the mass of the outer disk
is believed to be $\simlt 1-2\,M_{\rm Jup}$, as attested by the low mass
of WTTS disks
\citep{2005ApJ...631.1134A, 2007ApJ...671.1800A, 2008ApJ...686L.115C, 2010ApJ...712..925C}.
During this ``disk dissipation" stage, the SEDs of the WTTS disks present  
a wide range of morphologies, as expected for disks with inner holes of
different sizes
\citep{2006ApJ...645.1283P, 2007ApJ...667..308C, 2010ApJ...724..835W}.
Once the remaining gas photoevaporates, the dynamics of the solid particles
become dominated by radiation effects (Figure~\ref{fig:evolution}d).  
Although the small ($r\simlt 1\,\mu$m) grains are quickly blown out by 
radiation pressure, slightly larger ones spiral in due to the
Poynting-Robertson effect and eventually evaporate when they reach
the dust sublimation radius.  What is left represents the initial
conditions of a debris disk: a gas poor disk with large grains,
planetesimals and/or planets. 
The fact that the vast majority of WTTS show no evidence for a disk
(primordial or debris) implies that not every primordial disk
evolves into a detectable debris disks.
Whether some disks remain detectable throughout the primordial
to debris disk transition or  there is always a quiescent  period 
between these two stages still remains to be established.

\subsubsection{The age variable}
\label{sec:disk_evolution_age}
Although it is true that there is some discernible correlation between the
ages of pre-main sequence stars and the evolutionary stages of their disks,
this correlation is rather weak.
In fact,  circumstellar disks in every stage of evolution, from massive
primordial disks to debris disks and completely dissipated disks,
are seen in stellar clusters and associations with ages ranging from
$\simlt 1$\,Myr to $\sim 10$\,Myr
\citep[e.g.,][]{2010ApJ...708.1107M}.
Diskless WTTS in the core of the young Ophiuchus molecular cloud and
the relatively old, gas-rich TW Hydra are good examples of these extremes.
This weak correlation between stellar age and disk evolutionary stage
can be explained by the combination of two factors:  the wide range in
the duration of the ``mass depletion'' stage and the short time scale
of the ``disk dissipation'' phase.
On the one hand, as discussed in \S\ref{sec:environment_binaries},
circumprimary disks in medium
separation binary systems are likely to have truncated disks with small
initial masses.  Because the ``mass depletion'' stage in such objects is
expected to be very short ($\simlt 0.3$\,Myr), their circumstellar
disks can easily go through all  the evolutionary stages described
above in less than 1\,Myr.  However, initially massive
circumstellar disks evolving in isolation can in principle remain
optically thick at all infrared wavelengths for up to $\sim 10$\,Myr.

\subsubsection{Evidence for alternative evolutionary paths}
\label{sec:disk_evolution_alternative}
The observed properties of most circumstellar disks are consistent
with the evolutionary sequence shown in Figure~\ref{fig:evolution};
however, some outliers do exist, showing that not all disks follow the same
sequence of event in an orderly manner.
Perhaps the most intriguing outliers are accreting objects that have
cleared out inner disks but retained massive ($> 10\,M_{\rm Jup}$)
outer disks.  Examples of such objects, which represent a small
subgroup of the so-called transition disks, include DM Tau, GM Aur
\citep{2007MNRAS.378..369N},
and RX~J1633.9-2442 \citep{2010ApJ...712..925C}.
Their large accretion rates, inner radii, and disk masses make their
holes incompatible with the evolutionary sequence discussed above.
We must then consider an additional agent, most likely dynamical clearing
by a (sub)stellar or planetary-mass companion.
The nature of these objects is discussed in some detail in the following
section.

\subsection{Section summary}   
\begin{itemize}
 \item Protoplanetary disks evolve through a variety of processes, including
       viscous transport, photoevaporation from the central star,
       grain growth and dust settling, and dynamical interaction with
       (sub)stellar and planetary-mass companions.
 \item Photoevaporative flows from disk surfaces have been observed
       but the models disagree on the relative importance of FUV, EUV,
       and X-ray photoevaporation.
 \item There is strong evidence for grain growth to millimeter
       (and, in some cases, centimeter) sizes but the presence and
       distribution of larger bodies remain unconstrained. 
 \item Most protostellar disks go through a slow ``mass depletion''
       phase followed by a rapid ``disk dissipation'' stage.
       As the accretion rate steadily drops below the photoevaporative rate,
       the disk is rapidly eroded from the inside-out.
       A wide range in the duration of the two phases, together with
       the intrinsic dispersion of disk masses and radii, weakens
       the correlation between stellar age and disk evolutionary stage.
\end{itemize}

\section{TRANSITION DISKS}
\label{sec:transition_disks}
Transition  disks were first identified by \emph{IRAS} as objects with little
or no excess emission at $\lambda< 10\,\mu$m and a significant excess at
$\lambda\geq 10\,\mu$m \citep{1989AJ.....97.1451S, 1996AJ....111.2066W}.
The lack of near-infrared excess was interpreted as a diagnostic of inner
disk clearing possibly connected to the early stages of planet formation.
Because of this possible connection, transition disks have received special
attention in circumstellar disks studies even though they represent a
small percentage of the disk population in nearby star-forming regions. 
In this section we discuss the diversity of SED morphologies presented
by transition disks, their incidence, and their connection to planet
formation and other disk evolution processes.

\subsection{SED diversity and interpretation}
\label{sec:transition_disks_diversity}
\begin{figure}
\centerline{\psfig{figure=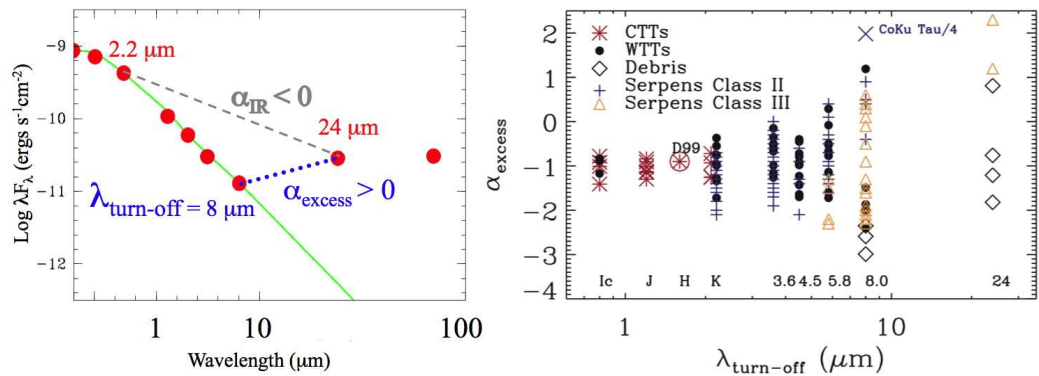,width=5.5in}}
\caption{
(a) A schematic of the $\alpha_{\rm excess}$ and $\lambda_{\rm turn-off}$
classification for a transition disk.
Although the $\alpha_{\rm IR}$  value is typical of a CTTS with a full disk,
the long $\lambda_{\rm turn-off}$ wavelength and positive
$\alpha_{\rm excess}$ indicate the presence of an inner hole. 
(b) Distribution of $\alpha_{\rm excess}$ with respect to
$\lambda_{\rm turn-off}$ for a range of different disk
evolutionary types \citep{2007ApJ...663.1149H}.
The diagram demonstrates the increasing range of possible
$\alpha_{\rm excess}$ values at longer $\lambda_{\rm turn-off}$ wavelengths
due to the diversity of transition disk SED morphologies.
}
\label{fig:lambda_alpha}
\end{figure}

Transition disks present a variety of infrared SED morphologies and their
diversity is not properly captured by the traditional classification scheme
of young stellar objects (i.e., the Class I through III system,
see \S~\ref{sec:classification}).
In order to better describe  the shape of transition disk SEDs,
\cite{2007ApJ...667..308C} introduced a two-parameter scheme based on the
longest wavelength at which the observed flux is dominated by the
stellar photosphere, $\lambda_{\rm turn-off}$, and the slope of the
infrared excess, $\alpha_{\rm excess}$, computed from $\lambda_{\rm turn-off}$
to $24\,\mu$m (Figure~\ref{fig:lambda_alpha}a.)

Because of their diversity, the precise definition of what constitutes
a transition object found in the disk-evolution literature is far from
homogeneous.  Transition disks have been defined as objects with no
detectable near-infrared excess, steep slopes in the mid-infrared, and large
far-infrared excesses \citep{2010ApJ...708.1107M, 2010ApJ...710..597S}.
This definition is the most restrictive one and effectively corresponds
to objects with $\lambda_{\rm turn-off}\simgt 4.5-8.0\,\mu$m and
$\alpha_{\rm excess} > 0$. 
The above definition has been relaxed by some researchers
(Brown et al. 2007;  Merin et al. 2010) to include objects with small,
but still detectable, near-infrared excesses. Transition disks have also been
more broadly defined in terms of a significant decrement relative to the
median SED of CTTS at any or all infrared wavelengths
\citep{2007MNRAS.378..369N, 2010ApJ...712..925C}.
Several nouns have recently emerged in the literature to distinct
sub-classes of transition disks \citep{2009arXiv0901.1691E}.
Objects with no detectable near-infrared excess and $\alpha_{\rm excess} > 0$
have been called ``classical" transition disks \citep{2010ApJ...708.1107M},
whereas other objects with a sharp rise in their mid-infrared SEDs
have  been termed ``cold disks" 
\citep{2007ApJ...664L.107B, 2010ApJ...718.1200M}
regardless of the presence of near-infrared excess. 
Disks with a significant flux decrement at all infrared wavelengths relative
to the SED of an optically thick disk extending to the dust sublimation
temperature typically have $\alpha_{\rm excess} < 0$
and been referred to as ``anemic'' disks \citep{2006AJ....131.1574L},
``homologously depleted'' disks \citep{2009ApJ...698....1C},
or ``weak excess'' disks \citep{2010ApJ...708.1107M}.
Finally, disks with evidence for an optically thin (to starlight)
gap separating optically thick inner and outer disk components have
been called ``pre-transition'' disks
\citep{2007ApJ...670L.135E, 2010ApJ...717..441E}
because they are believed to be precursors of objects with sharp, but
empty inner holes (i.e.,  the ``classical'' transition disks mentioned above).

The inherent diversity of transition disk SEDs is encapsulated in
Figure~\ref{fig:lambda_alpha}b which shows how the range of
slopes, $\alpha_{\rm excess}$, increases as the point at which
excess emission is detected, $\lambda_{\rm turn-off}$, increases.
Disks without inner holes, $\lambda_{\rm turn-off}\simlt 2\,\mu$m,
have similar mid-infrared SED slopes but this is not the case
for larger $\lambda_{\rm turn-off}$ suggesting a range of processes
by which the inner hole is formed and develops.

\begin{figure}
\centerline{\psfig{figure=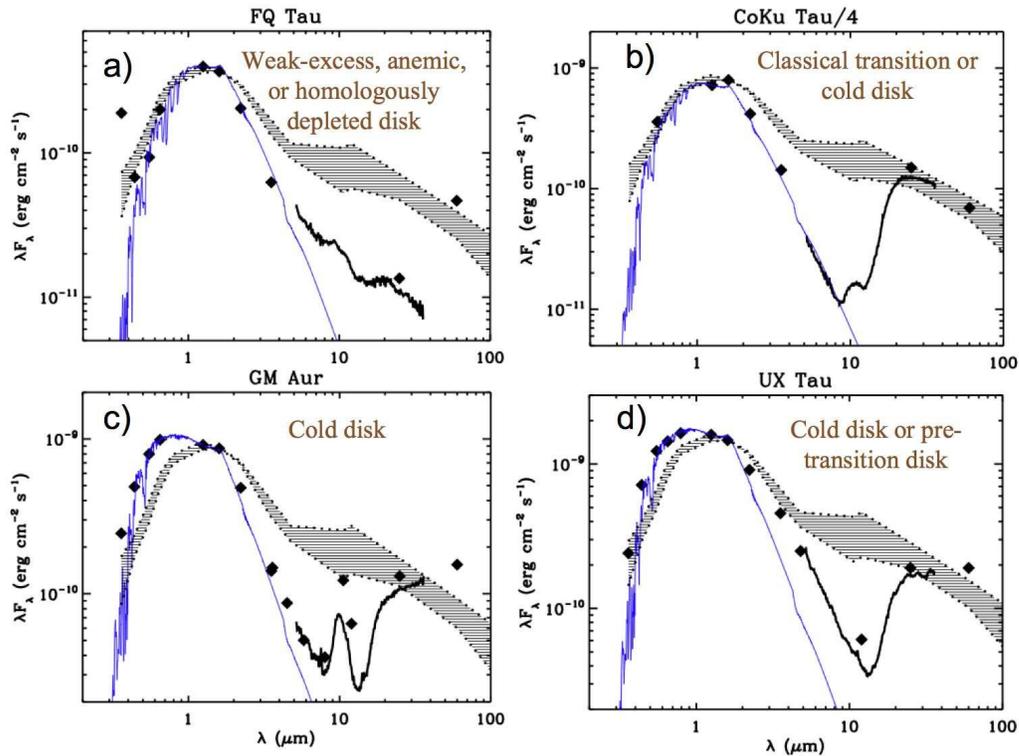,width=5.5in}}
\caption{
The diversity of transition disks SEDs.
The points represent photometry from optical to mid-infrared
wavelengths, whereas the dark solid lines are \emph{Spitzer}
infrared spectra. The stellar photosphere is shown as a blue curve,
and the dark hashed region shows the range of SEDs for typical
accreting T Tauri stars.
(a) A weak-excess, anemic, or homologously depleted disk has a
significant flux decrement at all mid-infrared wavelengths relative
to the T Tauri SED.
(b) A cold disk or ``classical" transition disk
displays excess emission above the photosphere only at
mid-infrared wavelengths and beyond.
(c) A cold disk with little near-infrared emission and a strong
$10\,\mu$m silicate feature.
(d) A cold disk with near-infrared excess emission. This can also
be considered a pre-transition disk because its SED can be modeled
with an optically thin gap separating optically thick inner and outer
disk components.
Figure adapted from \cite{2007MNRAS.378..369N}.
}
\label{fig:tran_types}
\end{figure}

Examples of the different types of transition disk SEDs are shown in
Figure~\ref{fig:tran_types}.
The SEDs that decrease in the mid-infrared, $\alpha_{\rm excess} \simlt 0$,
have generally been interpreted as evidence for grain growth and dust
settling toward the midplane
\citep{2006AJ....131.1574L, 2010ApJ...710..597S, 2010ApJ...712..925C}.
Both processes tend to reduce the mid-infrared fluxes.
Grain growth removes small grains and thus reduces the opacity of the
inner disk. Dust settling results in flatter disks that intercept and
reprocess a smaller fraction of the stellar radiation.
This group of objects includes anemic, homologously depleted disks,
and weak-excess disks.
Objects with rising mid-infrared SEDs, $\alpha_{\rm excess} > 0$,
a group that includes classical
transition disks, cold disks, and pre-transition disks, are more
consistent with a steep radial dependence in the dust opacity
resulting in a sharp boundary to the inner opacity hole.
These may be due to dynamical clearing by a companion,
possibly planetary \citep{1994ApJ...421..651A},
or photoevaporation \citep{2006MNRAS.369..229A}.
Debris disks also have sharp inner holes but they can be
distinguished by their low luminosities and lack of gas.

\subsection{Incidence} 
\label{sec:transition_disks_incidence}
Establishing the incidence of transition disks is not straightforward
given the different definitions used by different studies.
Also, samples are usually not complete and thus suffer from
selection effects. Furthermore, several background objects can mimic
the  SEDs of some types of  transition disks and skew the statistics.
In particular, Asymptotic Giant Branch stars and classical Be stars
can be easily confused with ``weak excess" transition disks
\citep{2010ApJ...714..778O, 2010ApJ...712..925C},
and the SEDs of edge-on protoplanetary
disks can look like those of cold disks \citep{2010ApJ...718.1200M}.
 
With these caveats, the fraction of disks in nearby star-forming regions
that are seen in a transition stage is thought to be $\simlt10-20$\%
\citep{2006AJ....131.1574L, 2009ApJ...698....1C, 2009ApJ...700.1017K,
2009A&A...504..461F, 2007ApJ...662.1067H, 2009AJ....137.4024D}.
Also, it is clear that ``weak-excess'' (or ``anemic'' and
``homologously depleted'') transition disks outnumber,
by factors of $\sim 2-3$, objects with sharp inner holes
\citep{2010ApJ...708.1107M, 2010ApJ...712..925C}.
The fraction of cold disks in nearby star-forming regions has been
recently established to lie between 5 and 10\%
\citep{2010ApJ...714..778O, 2010ApJ...718.1200M}.

Weak-excess disks seem to be more common in older clusters
\citep{2010ApJ...708.1107M}.
This is expected because disks should become flatter and lose mass with
time (see \S\ref{sec:disk_evolution_paths}).
Weak-excess disks appear to be much more abundant around young
(age $\sim 1$\,Myr) M-stars than around solar-type stars of the same age
\citep{2008ApJ...687.1145S}.
However, this may be a luminosity effect rather than an
evolutionary one.  Because M-type stars are much  fainter and cooler
than solar-type stars, they may present weak mid-infrared excess emission
even if the disk extends in to the dust sublimation radius
\citep{2009MNRAS.394L.141E}.
 
The relatively small number of objects seen in a transition stage
suggests that the evolutionary path through a transitional disk is
either uncommon or rapid.  However, observations show that an infrared excess
at a given wavelength is always accompanied by a larger excess
at longer wavelengths, out to $\sim 100\,\mu$m.
This implies that, unless some disks manage to lose the near-, mid-,
and far-infrared excess at exactly the same time, the near-infrared
excess always dissipates before the mid-infrared and far-infrared excess do.
No known process can remove the circumstellar dust at all radii
simultaneously, and even if grain growth or dynamical clearing
do not produce an inner hole, photoevaporation by the
central star will do so once the accretion rate through the disk
falls below the photoevaporation rate. 
Therefore, it is reasonable to conclude that transition disks represent
a common but relatively short phase in the evolution of a circumstellar
disk. Of course, how long the transition stage lasts for a given disk
strongly depends on the nature of  each object.
On the one hand, the transition phase should be very short, $<0.5$\,Myr,
if the inner hole is formed through photoevaporation.
However, an object could in principle show a transition disk
SED for a longer period of time if the inner opacity hole is due to
grain growth, or giant planet formation.

\subsection{Physical properties}
\label{sec:transition_disks_physical_properties}
Although it is clear that  many different processes can result in
transition disk SEDs, the relative importance of these processes is
not well understood.  Distinguishing among processes requires
additional observational constraints on the physical properties of
these objects, such as disk mass, accretion rates, fractional disk
luminosities ($L_{\rm disk}/L_{\rm star}$), and multiplicity information.
These observational constraints have only recently been obtained for
statistically significant samples of transition disks.
 
\cite{2007MNRAS.378..369N} investigated the SEDs, disk masses, and accretion
rates of over 60 Taurus pre-main sequence stars and identified
12 transition disks.
Based on their SEDs, 7 of these 12 could be further classified as weak
excess disks, 3 as classical transition disks, and 2 as pre-transition disks.
They found that the transition disks in their sample have larger average
disk masses and lower average accretion rates than non-transition disks
around single stars.
They concluded that, with the exception of CoKu Tau 4,
the SEDs were all more consistent with the giant planet formation
scenario than with photoevaporation models
(which at the time predicted negligible accretion rates and
lower disks masses) and grain growth (which would favor
higher accretion rates for a given disk mass).
However, because their sample was drawn from the \emph{Spitzer}
spectroscopic survey of Taurus presented by \cite{2006ApJS..165..568F},
who in turn selected their targets based on mid-infrared colors from
\emph{IRAS}, their sample was biased towards the brightest
objects in the mid-infrared and are unlikely to represent the overall
population of transition disks.

As a counter example, \cite{2008ApJ...686L.115C} studied over 20 WTTS with
transition disk SEDs, found that all had very small disk masses
($\simlt 2\,M_{\rm Jup}$), and concluded they were all consistent
with the EUV photoevaporation models of
\cite{2006MNRAS.369..216A, 2006MNRAS.369..229A}.
\cite{2010ApJ...710..597S} studied accretion rates in a sample of 95
members of the $\sim 4$\,Myr Tr 37 cluster. They found that half of
the 20 classical transition disks in their sample had evidence
for accretion (i.e. their holes contain gas but no dust) and half
were non-accreting  (i.e., their holes are really empty).
Furthermore, the accretors had rates that
were indistinguishable from those of regular CTTS in the cluster,
a result that is at odds with the Taurus sample studied by
\cite{2007MNRAS.378..369N}.
These discrepancies show that transition disks are a highly
heterogeneous group of objects and that the  mean properties of a
given sample are highly dependent on the details of the sample selection
criteria and, thus, should be interpreted with caution.

\begin{figure}
\centerline{\psfig{figure=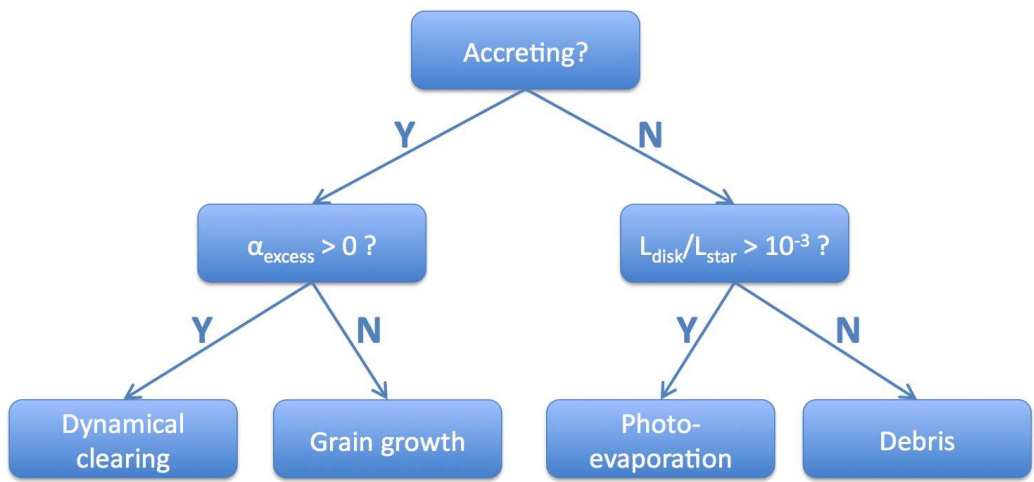,width=5.5in}}
\caption{
Decision tree to determine the dominant physical process for transition
disks, defined as having a significant flux decrement relative to
the median SED of CTTS at any or all infrared wavelengths.
This appears to be an exhaustive set of possibilities for
transition disks in Ophiuchus \citep{2010ApJ...712..925C}
but there may be rare additional possibilities in other regions.
}
\label{fig:tran_tree}
\end{figure}

\cite{2010ApJ...712..925C} studied a sample of 26 transition disks in Ophiuchus.
With the exception of edge-on disks, highly embedded objects, and the
lowest mass transition objects, all of which are too faint in the optical
to pass one of their selection criteria, the sample is likely to be
representative of the entire transition disk population in the Ophiuchus
molecular cloud.  They find that 9 of the 26 targets have low disk
masses ($< 2.5\,M_{\rm Jup}$) and negligible accretion,
and are thus consistent with photoevaporation.
4 of these 9 non-accreting objects have fractional
disk luminosities $< 10^{-3}$, however,
and could already be in a debris disk stage.
The other 17 objects in the sample are accreting.
13 of these are consistent with
dust settling and grain growth ($\alpha_{\rm excess} \simlt 0$),
whereas the other 4 have rising mid-infrared SEDs ($\alpha_{\rm excess} > 0$)
indicative of sharp inner holes, and are candidates for harboring
embedded giant planets.
A decision tree showing the different possibilities is shown in
Figure~\ref{fig:tran_tree}.

Interestingly, non-accreting objects with relatively massive disks
($\simgt 2.5\,M_{\rm Jup}$) are absent in the Ophiuchus transition
disk sample.  Such objects are rare but do exist. The ``cold disk" around
T~Cha is an example \citep{2007ApJ...664L.107B},
and its properties could be explained by a companion massive enough to 
disrupt accretion onto the star.
A few accreting binary systems are known, however,
for example, DQ~Tau \citep{2001ApJ...551..454C}
and CS~Cha \citep{2007ApJ...670L.135E},
which implies that a massive companion does not always stop accretion. 
The eccentricity of the orbit, and the viscosity and scale height
of the disk also determine whether accretion onto the star can continue
\citep{1996ApJ...467L..77A}.

Binarity is also known to result in transition disk SEDs.
Identifying close companions is particularly difficult, however,
especially if they are within the inner $\sim 5-10$\,AU of the disk,
which is beyond the near-infrared diffraction limit of $8-10$\,m class
telescopes at the distance of the nearest star-forming regions
($\sim 125-140$\,pc).    
Using the aperture masking technique in a single Keck telescope, 
\cite{2008ApJ...678L..59I} showed that CoKu Tau 4 is a near-equal mass
binary system, which explains the inner hole that had been inferred
from its SED; nevertheless, it is clear that not all sharp inner holes
are due to binarity.  DM Tau, GM Aur, LkCa 15, UX Tau, and RY Tau have
all been observed with the Keck interferometer \citep{2010ApJ...710..265P}.
For these objects,  all of which show evidence for accretion unlike
CoKu Tau 4, stellar companions with flux ratios $\simlt 20$ can be
ruled out down to sub-AU separations.

\subsection{Resolved observations}
\label{sec:transition_disks_resolved}
Although SED modeling provides strong evidence for the presence of inner
holes in some transition disks, this evidence is indirect and
model-dependent. Long baseline interferometry at (sub-)millimeter
wavelengths, however, can directly image the inner holes
of some transition disks in nearby star-forming regions.

\begin{figure}
\centerline{\psfig{figure=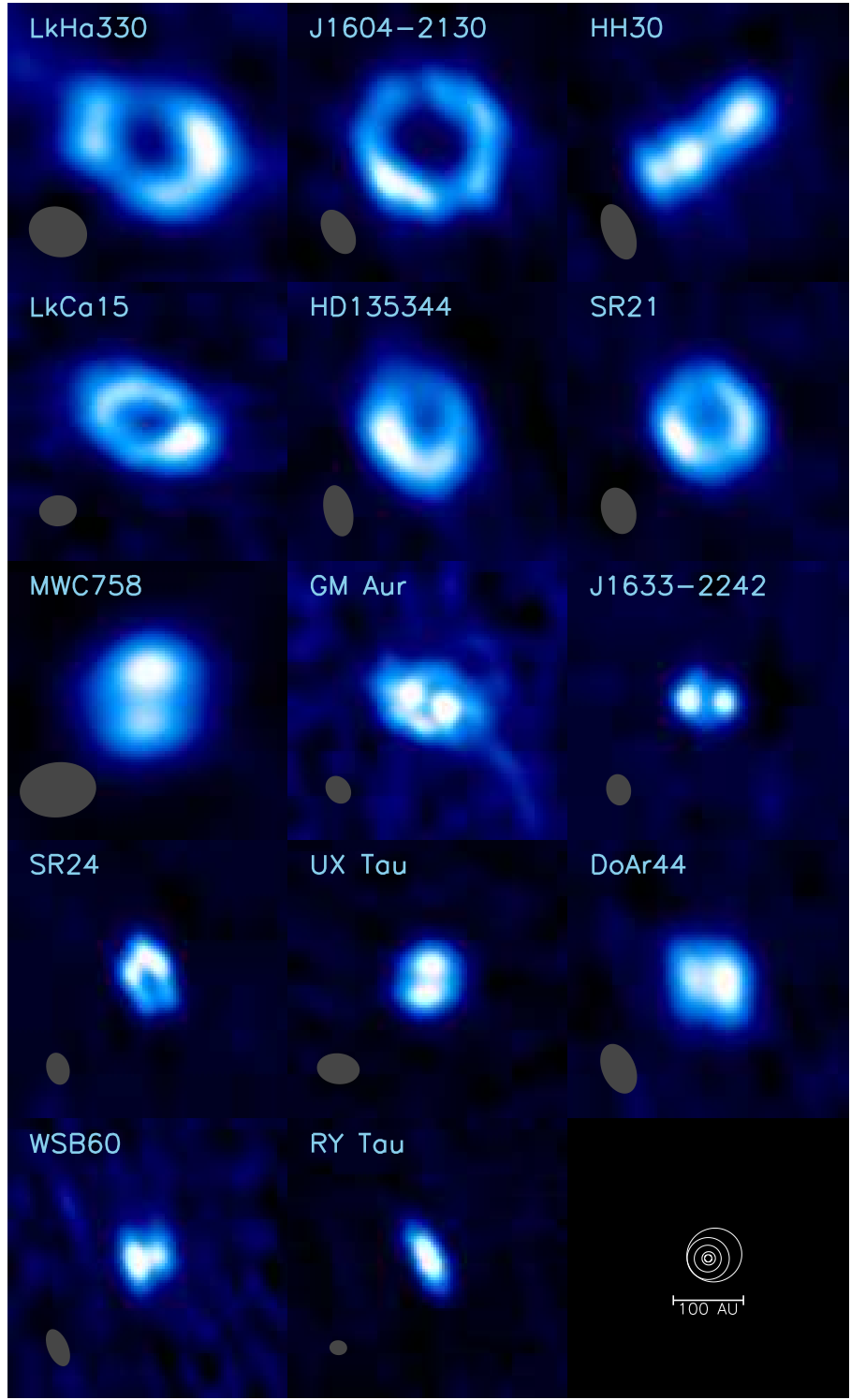,angle=0,width=4.7in}}
\caption{
Montage of (sub-)millimeter images of protoplanetary disks with resolved
inner holes.
Each panel is 400\,AU on a side, and the images are shown stretched
over 99\% of the range from minimum to maximum.
The gray ellipse at the lower left of each panel shows the synthesized
beam size, and ranges from $0\farcs 15$ to $0\farcs 8$.
The orbits of the giant planets and Pluto in the Solar System are shown
in the lower right panel for scale.
The data are from a variety of published studies with the
\emph{SMA}, \emph{Plateau de Bure}, and \emph{CARMA} interferometers
\citep{2009ApJ...704..496B, 2009ApJ...701..260I, 2009ApJ...698..131H, 2010ApJ...723.1241A, 2006A&A...460L..43P, 2008A&A...478L..31G}
as well as not yet published results.
}
\label{fig:tran_images}
\end{figure}

Inner holes have finally been resolved in a number of transition disks
in recent years. Figure~\ref{fig:tran_images} displays a montage
of images at the same linear scale in comparison to the orbits
of the giant planets in the Solar System.
The holes can be clearly seen in the face-on disks and are apparent
through a central dip in the emission for edge-on disks.

Because the emission is optically thin in these (sub-)millimeter data,
the images directly show the mass surface density profile of the disk
and can be used to measure the size and sharpness of inner holes.
In most cases, the (sub-)millimeter interferometry confirms the
expected sizes of the inner holes (ranging from $\sim 4$\,AU to
$\sim 50$\,AU in radius) and their sharpness (i.e., a large increase
in the dust density over a small range in radii) predicted from the SED.
For example, \cite{2009ApJ...704..496B} and \cite{2010ApJ...723.1241A}
show that the flux decrement toward the centers of their imaged
transition disks are consistent with step functions with
dust depletion factors ranging from $10^2$ to $10^4$.

Several of these transition disks are gas rich.
In two cases, GM Aur and J1604-2130,
where the millimeter CO emission is resolved to sufficient
detail either spectroscopically or spatially, the central cavity
is found to be not only dust but also gas deficient
\citep[respectively]{2008A&A...490L..15D}.
However, \cite{2008ApJ...684.1323P}
found mid-infrared CO emission from well inside the dust gaps in
SR 21, HD 135344, and TW Hya.

The details in millimeter interferometric images may contain
dynamical hints of giant planets.
All the images in Figure~\ref{fig:tran_images} appear to be
asymmetric but these are no more than suggestions given the modest
signal-to-noise ratio and poor sampling of the Fourier plane.
\cite{2009ApJ...698..131H} note a warp between the dust and gas
structures in GM Aur.
High fidelity images with \emph{ALMA} will
soon allow us to study the structure of circumstellar disks in much
greater detail, and the incidence of radial gaps, density waves, warps,
and other possible azimuthal asymmetries should become evident. 
It may well be possible to find earlier stages in the
disk clearing process through a decline in the millimeter flux
before the infrared opacity drops below unity \citep{2008ApJ...678L.133A}.

In addition to the (sub-)millimeter images discussed above,
in at least one case, LkCa 15, the inner hole has been imaged in the
near-infrared from starlight reflected from the edge of a disk
wall $\sim 50$\,AU in radius \citep{2010ApJ...718L..87T}
once again confirming the structure inferred from modeling the
SED of the object.

\subsection{Section summary} 
\begin{itemize}     
 \item Transition disks can be broadly defined as disks with a significant
       flux decrement relative to the median SED of CTTS at any or all
       infrared wavelengths. They constitute at most 20\% of the disk
       population.
 \item There is a wide range of transition disk SEDs, indicative of
       the varied physical processes, photoevaporation,
       grain growth, and dynamical interactions with companions or
       planets, that produce them.
 \item Grain growth and dust settling produce SEDs with falling
       mid-infrared emission ($\alpha_{\rm excess}\simlt 0$).
       SEDs that rise in the mid-infrared ($\alpha_{\rm excess} > 0$)
       are more consistent with a sharp boundary to the inner
       hole due to photoevaporation or dynamical interactions.
 \item Giant planet formation best explains the combination of
       accretion and steeply rising mid-infrared emission in moderately
       massive disks, $M\simgt 3\,M_{\rm Jup}$.
\end{itemize}

\section{SUMMARY POINTS}
\label{sec:summary}
Observations of protoplanetary disks are challenging due to their
small size, low masses, and cool temperatures.
However a number of basic facts have been clearly established.
Mid-infrared observations of optically thick emission
provide the most sensitive measures of the presence of a disk through
which we infer their occurrence and lifetimes.
Millimeter wavelength observations of optically thin emission
provide the best measures of disk masses and, through interferometry,
resolved images of their structure.
The interpretation of the data is complicated by the uncertainties
in our knowledge of grain growth and settling, gas dispersal,
and the feedback between composition and structure.
Nevertheless, it is clear that there are a variety of evolutionary
pathways, and many different physical processes competing with each
other, including planet formation.
Even the restricted subject area of this review on observations of the
outer disk is so large that we have given a summary to each section.
Here, we distill these summaries yet further into six main points.

\begin{itemize}     
 \item Circumstellar disks form almost immediately after a
       molecular core collapses. They appear to be highly
       unstable at early times and accrete in bursts onto the
       central protostar.
 \item By the time a YSO becomes optically visible,
       the mass of the now protoplanetary disk averages 1\%
       that of the central star. Its surface density increases
       approximately inversely with radius and it has a soft,
       exponentially tapered edge between a few tens to hundreds of AU.
       The disks are now generally stable and rotate with a Keplerian
       velocity profile.
       About 15\% of disks around solar mass stars have a MMSN
       ($10\,M_{\rm Jup}$) of material within 50\,AU radius.
 \item Disks around solar and lower mass stars have a median lifetime
       between 2 and 3\,Myr but with a large dispersion from less than
       1\,Myr to a maximum of 10\,Myr.
       Lifetimes are shorter around higher mass stars and binaries
       with semi-major axes between 5 and 100\,AU.
       The large scale environment is relatively unimportant:
       disruption by stellar flybys is very rare and photoevaporation
       by massive stars generally only erodes the outer disk,
       beyond about 50\,AU.
 \item Protoplanetary disks evolve through a variety of processes, including
       viscous transport, photoevaporation by the central star,
       grain growth and dust settling, and dynamical interaction with
       (sub)stellar and planetary-mass companions.
       Most disks evolve via a slow decrease in the mass of gas and
       small particles followed by rapid disk clearing at all radii.
 \item The growth of dust grains from sub-micron sizes in the ISM
       to millimeter sizes in disks occurs early and continues
       with time. The presence of snowballs or pebbles several
       centimeters in size is hard to measure but has been inferred
       in a couple of cases.
 \item About 10-20\% of disks show mid-infrared dips indicative
       of inner holes. They present a wide range of SED morphologies
       and physical properties
       (disk masses, accretion rates, and inner hole sizes).
       These transition disks are excellent laboratories to study
       the disk clearing phase and the formation of planets.
\end{itemize}

\section{FUTURE ISSUES}
\label{sec:future}
Protoplanetary disks emit predominantly at long wavelengths where the
atmospheric background is high, and where high resolution requires
interferometry. By necessity, many current observations are of the
brightest disks but these may not be truly representative.
Systematic studies of the median disk population are about to becomre routine.
Ongoing observations with \emph{Herschel} 
at far-infrared wavelengths around the peak of their SED are just beginning
to produce new insights into disk structure, chemistry, and evolution.
Within the next few years, \emph{ALMA} will dramatically expand upon the
pathfinder work of present-day interferometers at (sub-)millimeter wavelengths.
Further down the line lies the prospect of sensitive, high resolution
observations in the mid-infrared with thirty-meter class telescopes,
as well as at centimeter and longer wavelengths with the Square Kilometer Array.
There is an enormous range of issues to investigate with these and
other instruments. Here, we list those that we find particularly compelling.

\begin{itemize}
\item
\emph{Disk formation:}
With the high sensitivity and imaging fidelity of \emph{ALMA},
it will be possible to map faint isotopic lines of dense gas
tracers and search for small rotationally supported structures
in the centers of molecular cores. Instabilities may be
revealed through spiral waves and other asymmetries and
can provide an independent dynamical disk mass estimate.
Optically thick lines seen in absorption against the disk
continuum will show the infall of material from the core.
Multi-wavelength continuum imaging will track the increase
in the grain size distribution through the process.

\item
\emph{Peering into the terrestrial planet zone:}
Very high resolution observations, $\simlt 0\farcs 1$, at wavelengths
beyond a millimeter can image optically thin dust emission and resolve
structures in the terrestrial planet-forming zone, $R\simlt 5$\,AU,
of the closest disks.
The longer the wavelength of the observation, the larger the size of
the dust grains that can be detected.
The first systematic studies at centimeter wavelengths are
just beginning with the extended \emph{VLA}.
The Square Kilometer Array will measure the distribution of rocks
and snowballs up to meter sizes.
Together with \emph{ALMA}, resolved images from sub-millimeter
to centimeter wavelengths will show the radial variation of
grain growth.

\item
\emph{Disk chemistry:}
The ability to survey many disks in many lines with \emph{ALMA}
will revolutionize the young field of disk chemistry.
As different species and transitions are excited in different
regions of a disk, such observations will enable a far more complete
picture of the gas disk structure to be developed.
Observations of H$_2^{18}$O will reveal the water context of
disks and constrain the location of the snowline in a statistically
meaningful sample.
The detection of other molecular isotopologues will allow isotopic
abundances (and radial gradients) to be measured and directly
compared to cosmochemical studies of meteorites.

\item 
\emph{What is the overall evolution of the gas-to-dust mass ratio
in circumstellar disks?} 
Although the evolution of the dust ($r\simlt 1$\,mm) content in disks
can be traced reasonably well by current (sub)millimeter observations,
our current understanding of the evolution of the gas is very limited.
Observations of gas tracers with \emph{Herschel}
(e.g., [O I] at 63.2 $\mu$m),
large infrared telescopes (e.g., H$_{2}$ at 12.4 and 17.0 $\mu$m),
and \emph{ALMA} (e.g., rotational lines of CO and its isotopologues)
will reveal the evolution of the gas and the gas-to-dust ratio,
which is critical to understanding the formation of both terrestrial
and giant planets. 

\item
\emph{How and when do giant planets form?} 
Although much progress has been made in understanding 
the structure and evolution of circumstellar disks,
this fundamental question still remains unanswered. 
Detailed studies of embedded (Class I) disks with \emph{ALMA}
will help to establish whether massive young disks can be conducive to 
the formation of giant planets through gravitational instability. 
Similarly, planet searches in Class II YSOs, and transition disks
in particular, with thirty-meter class infrared telescopes
will identify young giant planets at the last stages of the
core accretion process. 
 
\item
\emph{Toward comprehensive disk evolution models:} 
To date most disk evolution models have focused on one or two
physical processes at a time (e.g., viscous accretion and
photoevaporation) while ignoring other equality important ones
(e.g., grain growth and dust settling and dynamical interactions
with (sub)stellar companions and/or young planets). 
In reality, however, it is clear that all these processes are 
likely to operate simultaneously and affect one another,
and that any realistic disk evolution model should include 
all known disk evolution mechanisms. 

\item
\emph{Placing our Solar System in context:}
The small number of protoplanetary disks that have been studied in detail
each have their own idiosyncracies.
Whereas there are rough matches to the mass, size, and surface density
profile of the MMSN, it is not clear how common these conditions were.
Improvements in technology at all wavelengths will allow more refined studies
of the closest disks and large surveys of more distant star-forming regions.
In tandem with the ever increasing knowledge-base on the number,
mass, and density distribution of exoplanets,
we will gather the detailed statistics necessary to understand what is
typical and what is atypical about the protosolar nebula.

\end{itemize}

\section{Acknowledgements}
We gratefully acknowledge Ewine van Dishoeck, Michiel Hogerheijde and
Neal Evans for their detailed comments. We also thank
Sean Andrews, Joanna Brown, Stephane Guilloteau, Paul Harvey,
Meredith Hughes, Andrea Isella, Antonella Natta, Ilaria Pascucci,
Charlie Qi, and Goran Sandell for reading an early version of this
manuscript, assistance with figures, or sharing data.
This work is funded through grants from the National Science Foundation
and the National Aeronautics and Space Administration
through the \emph{Spitzer} and \emph{Sagan} Fellowship programs.

\bibliographystyle{Astronomy}
\bibliography{references}


\begin{table}
\label{table1}
\caption{Classification of Young Stellar Objects}
\begin{tabular}{@{}cccl@{}}
\toprule
Class &           SED slope         & Physical properties & Observational characteristics \\
\colrule
  0   &              --             & $M_{\rm env} > M_{\rm star} > M_{\rm disk}$             & no optical or near-infrared emission \\
  I   & $\alpha_{\rm IR}>0.3$       & $M_{\rm star} > M_{\rm env}\sim M_{\rm disk}$           & generally optically obscured \\
 FS   & $-0.3<\alpha_{\rm IR}<0.3$  &                                                         & intermediate between Class I and II\\
 II   & $-1.6<\alpha_{\rm IR}<-0.3$ & $M_{\rm disk}/M_{\rm star}\sim 1\%, M_{\rm env}\sim 0 $ & accreting disk; strong H$\alpha$ and UV \\
III   & $\alpha_{\rm IR}<-1.6$      & $M_{\rm disk}/M_{\rm star}\ll 1\%,  M_{\rm env}\sim 0 $ & passive disk; no or very weak accretion \\
\botrule
\end{tabular}
\end{table}

\end{document}